\documentclass[aps,twocolumn,groupedaddress,showpacs]{revtex4}

\usepackage[T1]{fontenc}
\usepackage{graphicx}
\usepackage{epstopdf}
\usepackage{amsmath}
\usepackage{bm}
\usepackage{float}
\usepackage[caption=false]{subfig}
\usepackage{adjustbox}
\usepackage[pdftex,hyperfigures,breaklinks,colorlinks,citecolor=blue,linkcolor=red]{hyperref} 
\usepackage{multirow}
\usepackage{gensymb}
\graphicspath{{Figures/}}

\renewcommand{\figurename}{{\bf Figure}}

\begin{document}
\title{Monolingual and bilingual language networks in healthy subjects using functional MRI and graph theory}

 \author{Qiongge Li$^{1,2}$, Luca Pasquini$^{3,4}$, Gino Del Ferraro$^{1,3}$, Madeleine Gene$^{3}$, Kyung K. Peck$^{3,5}$, Hern\'an A. Makse$^{1}$ and Andrei I. Holodny$^{3,6,7}$}
 
\date{\today}

\affiliation{ $^{1}$ Levich Institute and Physics Department, City College of New York, New York, NY 10031}

\affiliation{$^{2}$Physics Department, Graduate Center of City University of New York, New York, NY 10016, USA}

\affiliation{$^{3}$Department of Radiology, Memorial Sloan Kettering Cancer Center, New York, NY 10065, USA}   

\affiliation{$^{4}$Neuroradiology Unit, NESMOS Department, Sant'Andrea Hospital, La Sapienza University, Rome, RM 00189, Italy}

\affiliation{$^{5}$Department of Medical Physics, Memorial Sloan Kettering Cancer Center, New York, NY 10065, USA}

\affiliation{$^{6}$New York University School of Medicine, New York, NY 10016, USA}
\affiliation{$^{7}$Neuroscience, Weill Medical College of Cornell University, New York, NY 10065, USA}

\begin{abstract}
\begin{center}
\textbf{\abstractname}
\end{center}

Pre-surgical language mapping with functional magnetic resonance imaging (fMRI) is routinely conducted to assist the neurosurgeon in preventing damage to brain regions responsible for language. Functional differences exist between the monolingual versus the bilingual brain, whereas clinical fMRI tasks are typically conducted in a single language. The presence of secondary language processing mechanisms is a potential source of error in the inferred language map. From fMRI data of healthy bilingual and monolingual subjects we obtain language maps as functional networks. Our results show a sub-network ``core'' architecture consisting of the Broca's, pre-supplementary motor, and premotor areas present across all subjects. Wernicke's Area was found to connect to the ``core'' to a different extent across groups. The $k$ core centrality measure shows ``core'' areas belong to the maximum core while WA and other fROIs vary across groups. The results may provide a benchmark to preserve equal treatment outcomes for bilingual patients.

\end{abstract}

\maketitle

\section{Introduction}\label{sec:introduction}

A functional language network (FLN) is a network model of interacting brain areas which are sensitive to language processing. Prior to brain surgery, an individual FLN is often constructed from language task functional magnetic resonance imaging (fMRI) data obtained from a pre-surgical language mapping routine. To the greatest extent possible, it is desirable to avoid damage to the language function since it is a critical function of the human brain. The identification of FLN thus serves two clinical purposes. First, it may assist the neurosurgeon to avoid further damage to critical, language-sensitive connections and regions. Second, a comparison of the map to an established benchmark in healthy individuals may indicate the specific components of the language network that have already been compromised by the pathological condition \cite{aubert2002modeling,li2020core}.

Recently, a white paper from the American Society of Functional Neuroradiology established the gold standard protocol for language pre-operative assessment, which includes sentence completion, silent word generation and a third variable task \cite{black2017american} for adults. The combination of sentence completion and silent word generation tasks has been described optimal for language localization and lateralization \cite{zaca2013role}, due to the combination of language comprehension and language production. 

Visually administered silently-generated language tasks should activate language areas related to speech comprehension and production through covert speech, relying on semantic and syntactic mental representations, which require word retrieval and articulatory planning \cite{price2012review, price2010anatomy,cappa2012imaging,bouchard2013functional}. Particularly, the silent word generation task, considered a phonemic fluency task, requires phonologic access, verbal working memory, and lexical search activity, which grant a strong activation and lateralization of frontal areas \cite{li2017lexical,emch2019neural,corrivetti2019dissociating}. Consequently, the task showed the most robust language localization and the most effective language lateralization in the frontal gyri (IFG, MFG, and SFG) of the dominant language hemisphere \cite{zaca2013role}. with optimal language localization \cite{zaca2013role}, being considered among the first choices in the state-of-the-art fMRI paradigm for clinical applications \cite{black2017american}.

In a previous study \cite{li2020core}, we established a functional language ``core'' sub-network as such a benchmark by analyzing 20 healthy subjects without regard to their mono- or multi-lingual status. However, there are known functional differences in the language network of bilinguals compared to monolinguals which may affect surgical management\cite{wong2016neurolinguistics}. 

Bilingualism requires the control of different language systems. It has been postulated that different languages are coexisting in the awareness of the bilingual subject and are in conflict for language production until a specific one is chosen \cite{wong2016neurolinguistics}. The consequence is an enhancement of the ``control'' areas in the brain which include the dorsolateral prefrontal cortex (DLPFC), the anterior cingulate cortex (CC), the basal ganglia (BG), and especially the left caudate nucleus in the case of bilingual brain \cite{seo2018bilingual}.

The activation of the DLPFC is deemed maximal when the subject is asked to switch between the known languages \cite{seo2018bilingual} and the BG may actively mediate signaling to the prefrontal cortex according to the changing target language being used by the subject at any given time. Specifically, the caudate may regulate the selection of the less accessible language \cite{seo2018bilingual,li2016functional}.

Although the particular signature of the bilingual language network has been widely investigated in the literature, its effect on clinical practice is still unclear. Particularly, the common language network of bilinguals in clinically-acquired phonemic fluency tasks may differ from that of monolinguals in both the active areas and the connectivity between active nodes. These differences may impact clinical practice such as pre-surgical planning in evaluating the relevance of each language area. In fact, most of the studies concerning the bilingual language network have aimed to investigate specific areas for bilingualism by employing ad-hoc fMRI tasks \cite{seo2018bilingual,li2016functional} that have a limited role in the clinical practice. There
is a gap with respect to investigating coordinated patterns of neural response in bilingual
individuals while using the language tasks actually employed by clinicians (e.g., for pre-surgical
planning).

The clinical task our subjects performed in this study is recognized as one of the standards in clinical practice \cite{brennan2007object,ramsey2001combined} and is routinely employed for pre-surgical MRI evaluation in patients with brain tumors. We provided the detailed information of this task in Sec. \ref{sec:functional_task}.

Language differences are a well-known limitation to
fMRI evaluation in clinical practice. Especially in the
case of minorities speaking a different language from the
one employed in the task, the results of the examination
may be difficult to interpret. Availability of information such as active fMRI clusters interdependence may significantly improve the clinical care of minorities. The capability of pointing out a specific hierarchy of active clusters on fMRI maps, characterized by a dominant cluster whose integrity is necessary for the stability of the network \cite{del2018finding}, as well as crucial links between network nodes, appears particularly relevant in the bilingual brain, whose peculiar network organization may emerge from clinically-relevant tasks.

Our first objective is to determine if the results from clinically employed tasks (the ones which the surgeon would actually see in clinical practice) would differ between bilinguals and monolinguals. Secondly, we sought to study the network architecture of the FLN in each group (monolinguals, bilinguals speaking Spanish, and bilinguals speaking English) and characterize any differences arising from centrality measurements. Particularly, we sought to assess the $k$ core, which is emerging as an important topological measure of networks since it reveals a robust and highly connected sub-network, called the $k$ max core (as described in Sec. \ref{fig:kcore_demo}) \cite{dorogovtsev2006k,morone2019k}; $k$ core has previously been employed to measure the stability of the most resilient functional structures in the brain \cite{li2020core,lucini2019brain} and may provide useful insights in addition to the functional connectivity map. 

To this end, we analyzed fMRI scans from 16 healthy subjects: eight bilinguals and eight monolinguals. For every bilingual subject, we conducted two scans, where the language task was conducted in Spanish (L1) and in English (L2). For the monolinguals, we conducted each scan in English. We employ standard methods to construct voxel-level functional networks from thresholding the fMRI correlations \cite{bullmore2009complex,hermundstad2013structural,gallos2012small}. Functional regions of interest (fROIs) were then identified for every individual and every individual subject's network was transformed so that each fROI represents a single node. Connections were then defined between fROIs. Finally, we examined the $k$-core structure at the voxel level for each group.

 

\section{Materials and Methods}\label{sec:data}
		
\subsection{Subjects}

The study design was prospective. We recruited participants on a voluntary basis, who, provided written consent for participating in the study. 
The study was approved by the Institutional Review Board and was carried out according to the Declaration of Helsinki.
Sixteen self-reportedly healthy right-handed adult subjects (mean age = 42.37 years and Standard deviation = 8.92, nine males and seven females) without any neurological history were included. Among the 16 subjects, there were eight monolinguals (speaking only English) and eight bilinguals (speaking Spanish (L1) as their native language and English (L2) as their acquired second language). All bilinguals had professional-level fluency in speaking English.

\subsection{Functional MRI task}\label{sec:functional_task}

For the fMRI task, all subjects performed a phonemic fluency letter task in response to task instructions delivered visually \cite{black2017american}. Each monolingual performed the task in English. Each bilingual performed the task in English and Spanish separately, resulting in two separate scans for each bilingual subject. We interchanged the order of English and Spanish tasks randomly. In the final data cohort, we had 24 task-based fMRI (tb-fMRI) scans, eight English scans from the monolingual subjects, and eight English scans plus eight Spanish scans from the bilingual subjects.

In the phonemic fluency task (letter task), subjects were asked to silently generate words that began with the letter (for example, given the letter ``B'', subjects would generate words such as ``BIRD'', ``BIKE'', ``BANK'', etc.). Subjects silently generated words without vocalization to avoid creating artifacts from jaw movement. Stimuli were displayed on a screen over eight stimulation epochs with each epoch lasting 20 sec.
During the task, two letters were presented in each stimulation epoch. Each epoch also consisted of a 30 sec resting period during which subjects were asked to focus on a blinking crosshair. Brain activity and head motion were monitored using Brainwave software (GE, Brainwave RT, Medical Numerics, Germantown, MD) allowing for real-time observation.

\subsection{Data acquisition}
A GE 3T scanner (750W, Milwaukee, Wisconsin, USA) and a 24-channel neurovascular head coil was employed to acquire the MR images. Functional images covering the whole brain were acquired using a $T_{2}^{*}$-weighted gradient echo echo-planar imaging sequence (repetition time ($TR$)) divided by (echo time ($TE$)) = 2500 ms/30 ms; slice thickness = 4 mm; matrix = $64\times64$; FOV = 240 mm; flip angle FA = $80 \degree$; voxel resolution = 4mm $\times$ 4mm $\times$ 4mm. In addition, functional coverage matching $T_{1}$-weighted 3D BRAVO (spoiled gradient recalled echo (SPGR) with inversion activated) images ($TR/TE$ = 8.2 ms /3.1 ms; slice thickness = 1 mm; Inversion Time = 450 ms; matrix = $240\times 240$, FA = $12 \degree$) were acquired for co-registration and deformation purposes. 

\subsection{Data pre-processing}

fMRI data were processed and analyzed using the software program Analysis of Functional NeuroImages (AFNI) \cite{cox1996afni}. Head motion correction was performed using 3D rigid-body registration. Spatial smoothing was applied to improve the signal-to-noise ratio using a Gaussian filter with 4 mm full width of half maximum. Corrections for linear trend and high frequency noise were also applied. Signal changes over time were cross-correlated with a mathematical Gaussian model of the hemodynamic response to neural activation. Cross-correlation involved convolving the modeled waveform corresponding to the task performance block with all pixel time courses on a pixel-by-pixel basis to generate functional activity data. Functional activation maps were generated at a threshold of $p < 0.001$. To reduce false positive activity from large venous structures or head motion, voxels in which the standard deviation of the acquired time series exceeded 8\% of the mean signal intensity were set to zero. 

\subsection{Individual brain network construction}\label{sec:individual}

\begin{figure}[!htbp]
		\centering
		\includegraphics[width=0.48\textwidth]{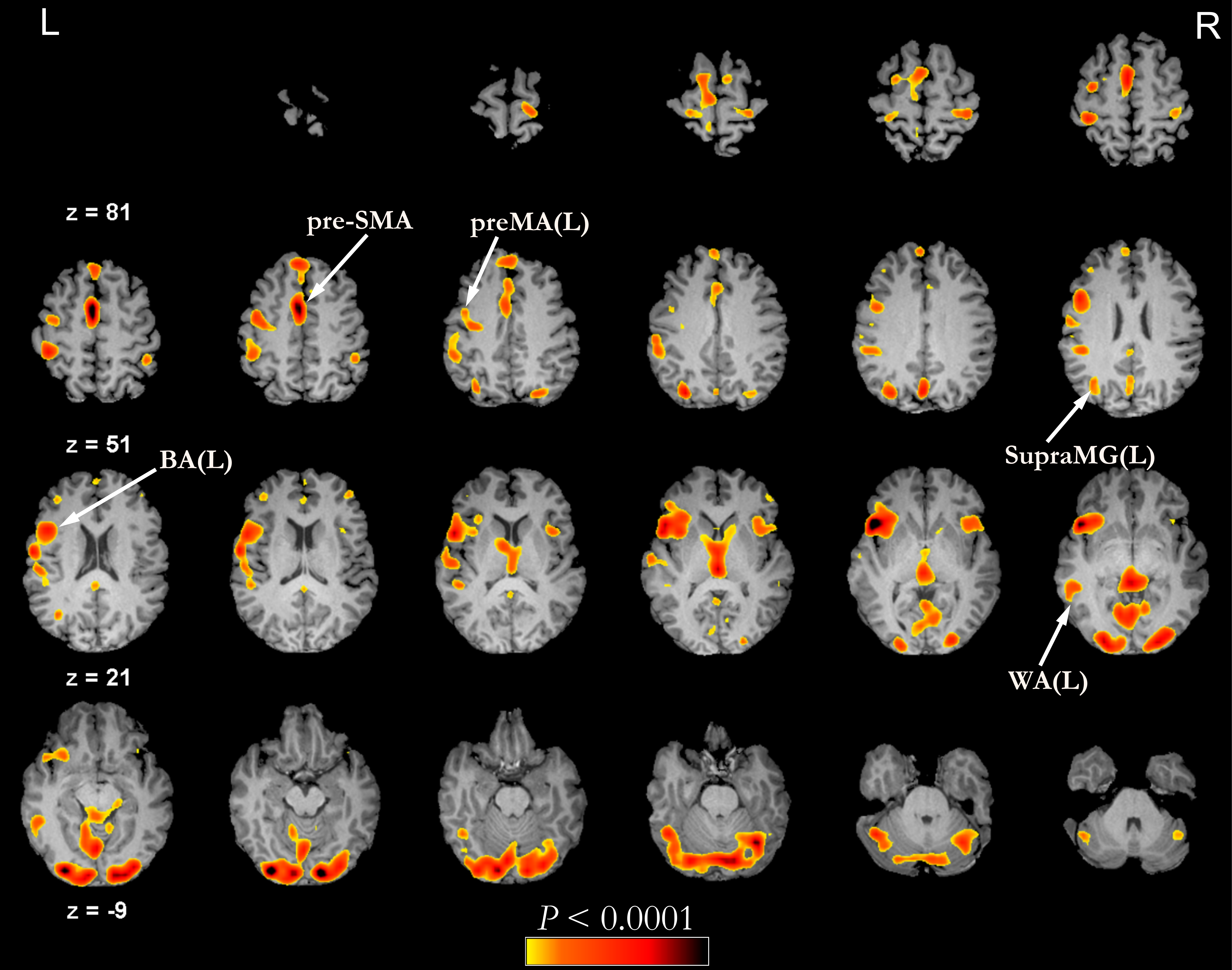}
		\caption{{\bf A representative subject's fMRI activation map overlaid on the anatomical MR image}. The reader's left-hand side is the subject's left hemisphere of the brain. The slice number is indicated by z. The areas highlighted in color correspond to fMRI active brain areas and the color bar at the bottom of the figure provide the $p$-values. Several regions have been labelled according to anatomical location. Areas such  as visual cortex which are unrelated to language but active during the task were not included. 3D Clusters were extracted and then named according to their anatomical locations such as left Broca's Area (BA(L)), left Wernicke's Area (WA(L)), pre-supplementary motor area (pre-SMA), left premotor area (preMA(L)) and left Supra-Marginal Gyrus (SupraMG(L)) marked on the image.}
		\label{fig:act_map}
	\end{figure}

Here, we briefly illustrate the method of construction of the functional network on two different scales(voxel and fROI), following a previous approach \cite{del2018finding,li2020core}. At the \textbf{voxel scale}, we first established the nodes of the functional languages network. Nodes were defined as ``activated'' voxels. To determine which voxels were activated, first, we extracted the pre-processed time series from all voxels of the functional image and then fitted each time-series to the transformed task model according to the general linear model's statistical approach \cite{friston1994statistical}. The voxels that passed the statistical significance test (we chose a small absolute threshold $p < 0.001$) were retained as activated voxels defining the network's nodes. A sample subject's resulting fMRI activation map is shown in Fig. \ref{fig:act_map}.

Because instructions were delivered visually, the visual cortex activates automatically. Furthermore, the subjects were permitted to keep their eyes open or closed during the task so it is customary to exclude the visual cortex from the fMRI activation map. These areas support non-linguistic processing and was therefore discarded from the analysis. These visual cortex regions were the only regions discarded from the analysis.

Then, we determined the modules or functional regions of interest (fROIs). An fROI was defined as a cluster of spatially proximate activated voxels (nodes). We identified separate clusters from the fMRI signal based on their anatomical proximity. The task of labeling of fROIs was performed by a neuroradiologist with extensive experience in clinical and research fMRI \cite{price2012review, price2010anatomy, benjamin2017presurgical,mandonnet2017surgical,matsuo2003discrimination,sarubbo2020mapping}.
	
	Although we used anatomical locations for labeling, we must clarify here that our module identifications were based on the functional signals rather than an anatomical atlas. There is high variance in the mapping of cognitive functions in individuals to the spatial location of anatomical landmarks; for this reason, we did not determine fROIs at the group level but rather at the individual level \cite{fedorenko2009neuroimaging}. Additionally, pre-operative fMRI results are always interpreted on an individual level \cite{peck2009presurgical}.  
	
	Having established the nodes and modules for the voxel scale network model, next, we defined the links between voxels (nodes) following standard methods of measuring statistical dependencies between activated voxels \cite{bullmore2009complex,hermundstad2013structural,gallos2012small,del2018finding,li2020core}. Links in the network were obtained by thresholding the voxel-voxel temporal correlation of the Blood-Oxygen-Level Dependent (BOLD) signal \cite{li2020core}, as shown below.:
	\begin{equation}
	C_{ij}=\frac{\langle x_i x_j \rangle - \langle x_i \rangle \langle x_j \rangle}{ \sqrt{(\langle x_i^2 \rangle -\langle x_i \rangle)^2 (\langle x_j^2 \rangle -\langle x_j \rangle)^2 } }.
	\end{equation}
	where $x_i$ is a vector encoding the fMRI time response of voxel $i$ and $\langle \cdot \rangle$ indicates a temporal average.

\begin{figure}[!htbp]

		\includegraphics[width=0.46\textwidth]{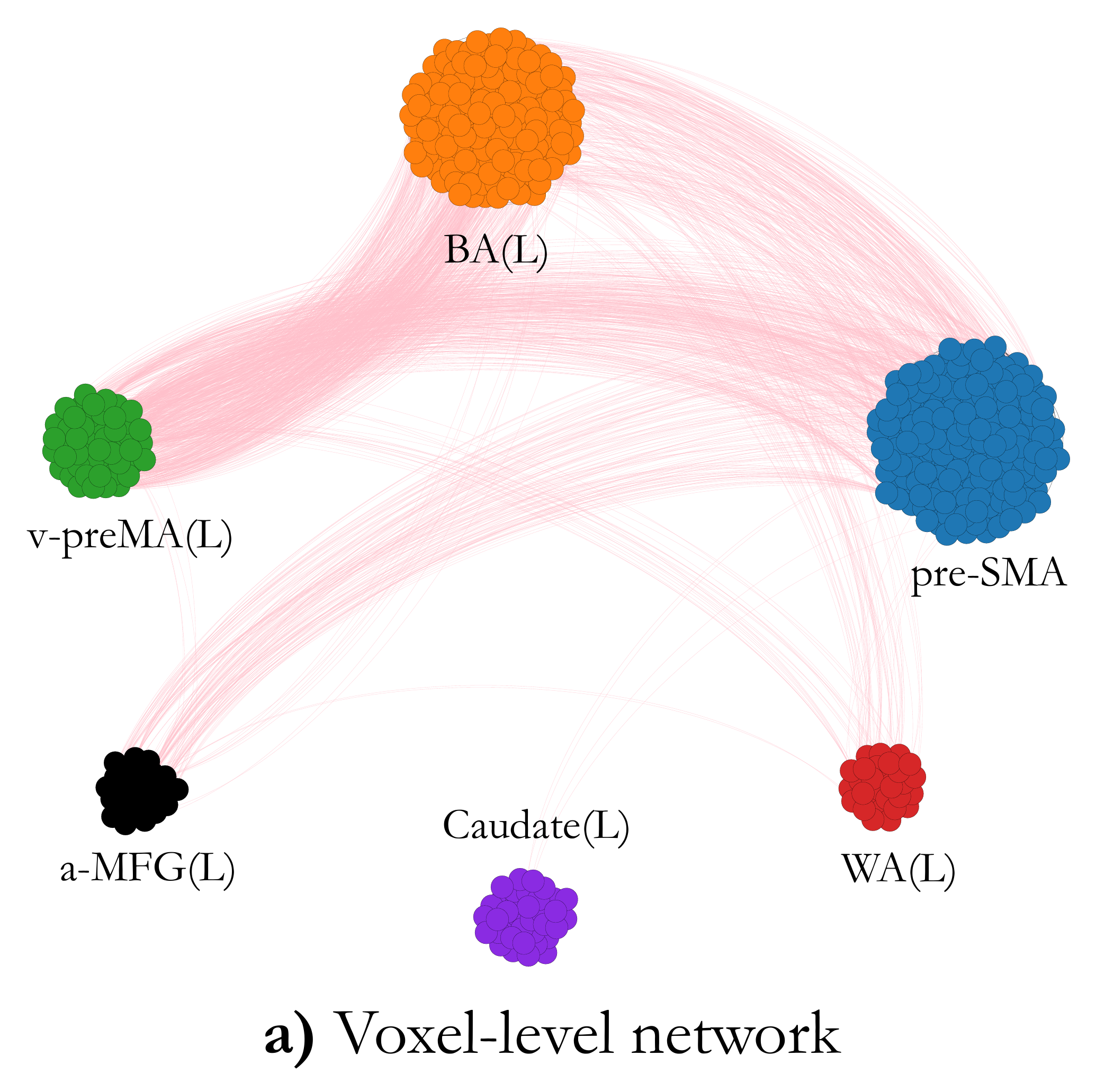}
		\includegraphics[width=0.43\textwidth]{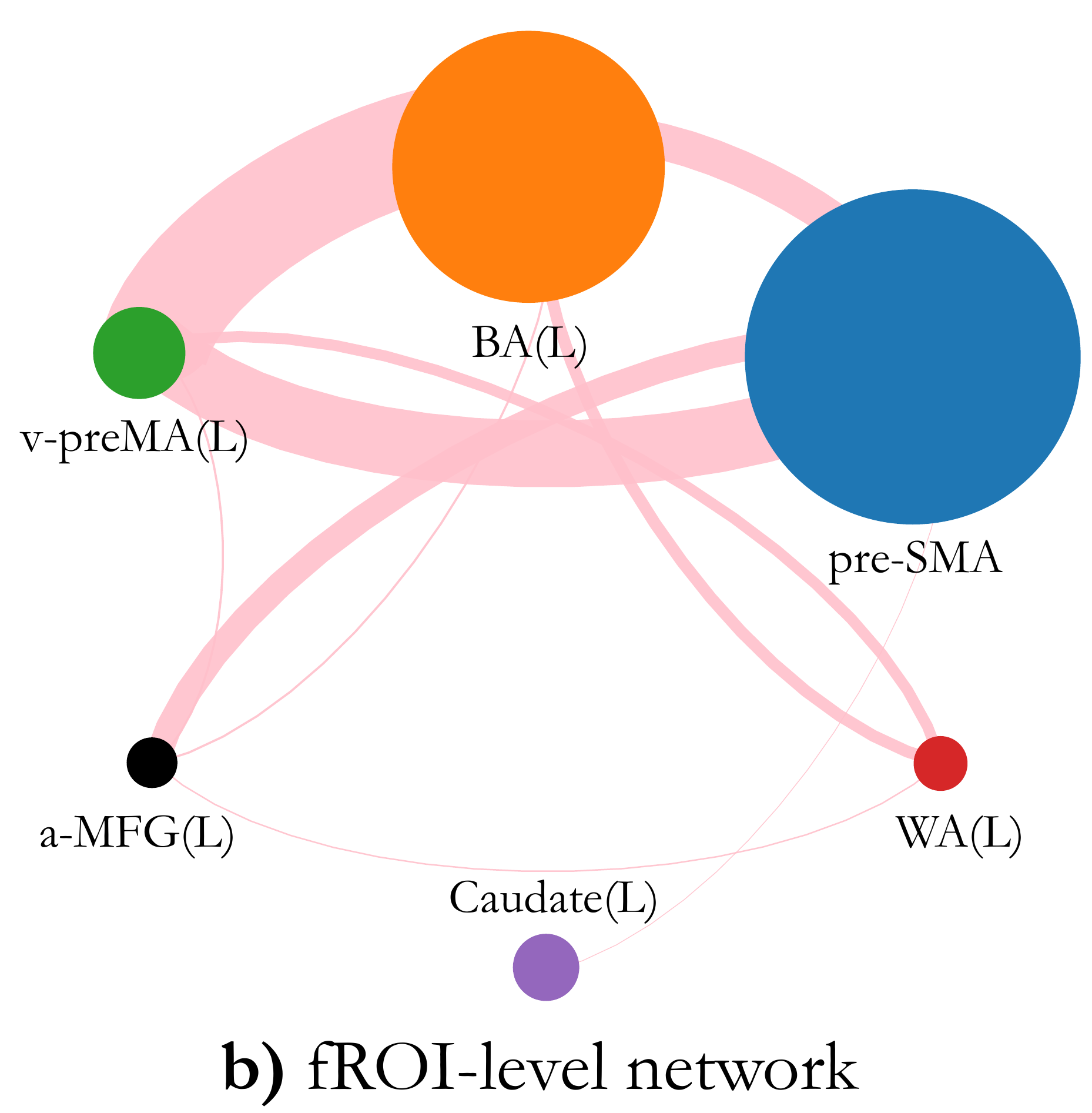}
		\caption{ A representative bilingual subject's network on the voxel scale (panel \textbf{a}) and fROI level (panel \textbf{b}). In panel \textbf{b} each node represents an fROI, and the node's size is proportional to the number of voxels in the fROI. Each link's thickness connecting two fROIs is proportional to the sum of all link weights inter-connecting the voxels between the two fROIs (as in Eq. \eqref{eq:W}). } 
		\label{fig:individual}
	\end{figure}

	A pair of voxels with a correlation above a certain threshold was considered as a link (absolute threshold). 
	In Fig.  \ref{fig:individual}, panel a), we display a realization of the voxel scale network for the sample subject in Fig. \ref{fig:act_map}. Each node represents a voxel, and nodes belonging to the same fROI are colored the same. The links connecting a pair of voxels belonging to different fROIs are shown as pink lines. The links connecting pairs of voxels within the same fROI are not shown. 

	In order to describe the FLN from a modular perspective, the network was transformed so that each fROI represents a single node. We refer to this as the \textbf{fROI scale} network. 
	To define the connectivity at the fROI-fROI scale, the functional link weight, $W_{ij}$, between two fROIs, labeled $i$ and $j$, is defined as the sum of all the binarized functional link weights, $w_{lm}$, connecting all pair of voxels, labeled $l$, and $m$, between the two fROIs, normalized by the sum of the two fROIs' sizes, $S_i$ and $S_j$,

	\begin{equation}
	W_{ij} = \frac{\sum_{l,m \in \{i \leftrightarrow j\}} w_{l,m}}{ (S_{i}+S_{j})}
	\label{eq:W}
	\end{equation} 
	
	Thus, there would be a non-zero fROI-fROI connection between any pair of regions such that a single voxel in each region is inter-connected. 
	
	We show a realization of an fROI-level network of the same representative subject in Fig. \ref{fig:individual} panel b). Here, each colored node represents a fROI. Each link's thickness connecting two fROIs is proportional to their link weight $W$. The apparent thickness of the fROI scale link relative to the voxel scale links may appear inconsistent but this can be explained by a visual artifact due to the geometrical arrangement of nodes in the voxel-scale diagram and by the normalization, Eq. \ref{eq:W}.

	We constructed both voxel and fROI scale networks from the tb-fMRI signal for each of the 24 individual scans of data partitioned into three groups: monolinguals, bilinguals speaking English, and bilinguals speaking Spanish, with each group containing eight networks.	Next, we measured the common network characteristics at the group level in order to estimate robust connectivity across subjects within a group.
	
\subsection{Common network construction across subjects}\label{sec:common}

We named the persistent functional architecture across subjects in a particular group at the fROI-level the `{\it common network}'. This common architecture was constructed for each group by retaining only those pairs of fROIs and those functional links connecting them that were present across all subjects, where we considered the number of appearances of the functional link within the group as a measure of frequency. 

The weight of the functional link connecting two fROIs, $i$ and $j$, in the common network ($W_{ij}^{\rm C}$) was defined as the average of the ${W}_{ij}$ connecting those fROIs across subjects:
\begin{equation} \label{eq:common_weights}
W_{ij}^{\rm C} = \frac 1 N \sum_{l=1}^N W_{ij}^{(l)}
\end{equation}
where $N$ is the number of individuals and where the link $W_{i,j}$ is nonzero.

\section{Results}
\subsection{Language proficiency tests}\label{sec:supp_proficiency}
Self-reported English and Spanish proficiency data were collected using two independent assessments: the four-item proficiency assessment \cite{klimidis2004brief} and the Language Experience and Proficiency Questionnaire (LEAP-Q) \cite{marian2007language}. 

For both assessments, bilinguals' English and Spanish proficiency scores were compared to one another using the Wilcoxon signed-rank test (paired). Bilinguals' English and Spanish proficiency scores were also individually compared to monolinguals' English proficiency scores using the Mann-Whitney $U$ test (unpaired).

No significant differences were found between bilinguals' English proficiency and Spanish proficiency in any of the language domains (speaking, understanding, reading) or in overall proficiency ($p$ > 0.05). There were also no significant differences between monolinguals' English proficiency and bilinguals' English proficiency across all measures ($p$ > 0.05). The same result was found between monolingual's English proficiency and bilinguals' Spanish proficiency ($p$ > 0.05). These results show that there were no significant differences in self-reported English and Spanish proficiency for monolingual and bilingual participants. 

\subsection{Individual networks}\label{sec:individual_result}

From the 24 scans, 17 activated areas (or fROIs) were identified via the procedure described in Sec. \ref{sec:individual}. A summary of these activated areas and their frequency of activation by subject is shown in Supplementary Table \ref{table:activeareas}. Both hemispheres demonstrate activation; however, left hemisphere dominance is clearly observed, which is expected from an fMRI language task in right-handed subjects, since language brain activation is mostly concentrated in the left hemisphere in these individuals \cite{isaacs2006degree}. This is especially true for phonemic fluency tasks, which showed optimal lateralization compared to other tasks \cite{zaca2013role,li2017lexical}.

Although 17 activated areas are detected, not all the areas are activated in each subject due to inter-subject variability. We observe that most of the areas are activated in the monolingual group (16/17), followed by the bilingual speaking Spanish group (13/17) and the bilingual speaking English group (12/17). 

 The areas that are activated in all subjects and all groups were the pre-supplementary motor area (pre-SMA), Broca's Area (BA(L)), and ventrical premotor area (v-pre-MA(L)). Wernicke's Area (WA(L)) is activated in bilingual speaking Spanish group in all eight subjects. Thus, these regions are included in the corresponding common networks by default. The anterior Middle Frontal Gyrus (L) (a-MFG(L)) and the Supra-Marginal Gyrus(L) (SupraMG(L)) activate with significant frequency (between 50 to 75 percent) of subjects for all three groups. 

Individual link weights for the fROI scale network are reported in Tables \ref{table:individual-M} - \ref{table:individual-BS}. We observe that, overall, the preMA is the most connected area across subjects in terms of connectivity
weight (strength). Only the shared links between subsets of activated fROIs are shown in these tables, with the subsets representing the core similarities between the groups as to be included in the common networks. As a general trend we can see that the strongest link (the link with the largest connectivity weight) is between v-preMA(L) and BA(L), followed by v-preMA(L) and pre-SMA, and then by pre-SMA and BA(L). 

We note several relatively small fROI level link weights such as $W_{ij} =0.01$ (for example refer to subject 6's connection between BA and WA in table \ref{table:individual-M}) in some subjects. These small values arise because the fROI scale network normalizes the link weights by the fROI sizes (number of voxels in the clusters), which can lead to apparently small link weights in some cases when the fROI size is large compared to relatively sparse interconnections.
 
\subsection{Common networks}\label{sec:common_result}

\begin{figure}[!htbp]
	\includegraphics[width=0.45\textwidth]{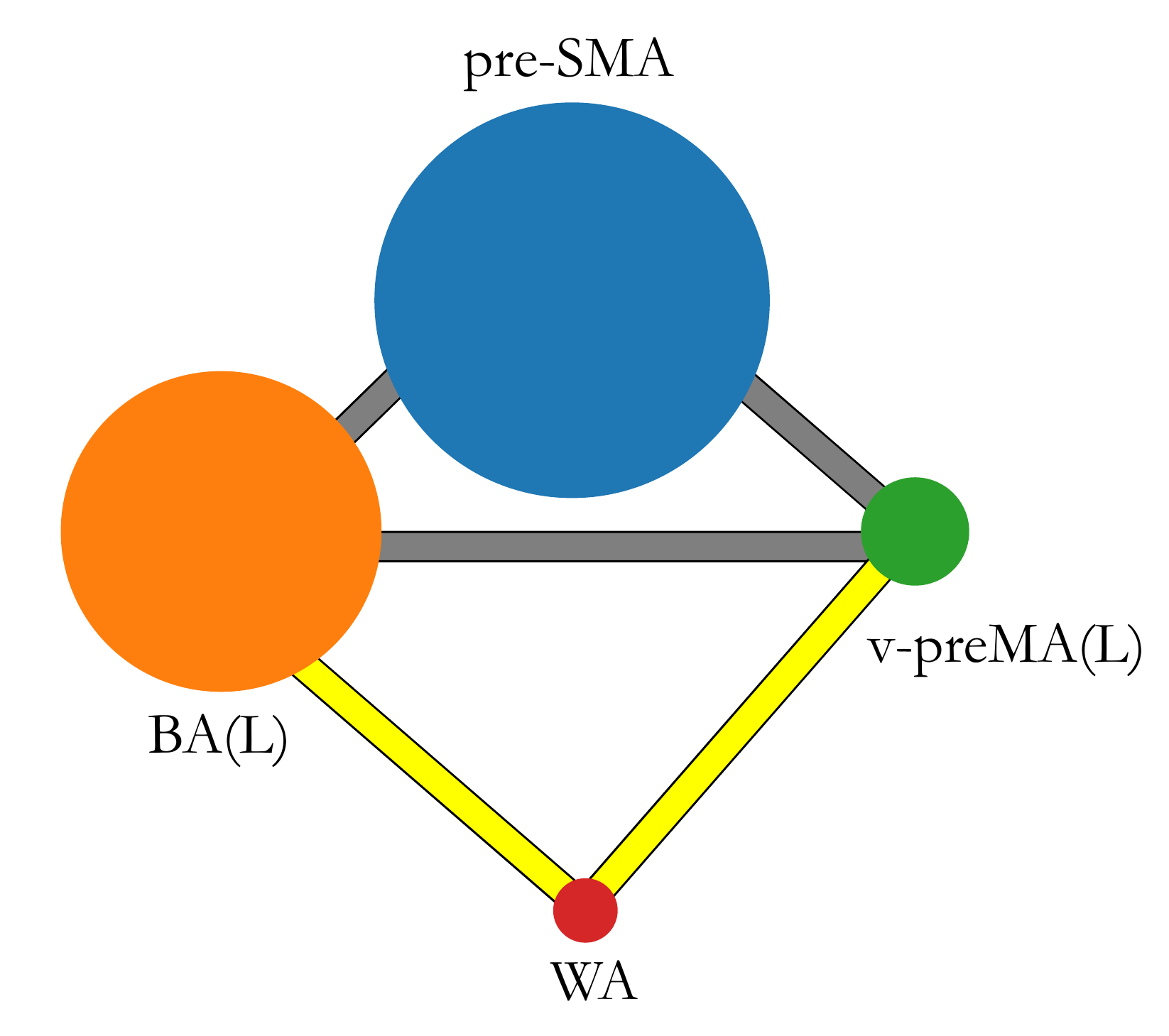}
	\caption{{\bf Common network structure in fROI level}. The colored nodes represent fROIs and their size is proportional to the averaged size of fROIs across all subjects. The solid gray links connecting pre-SMA, BA(L), and v-preMA(L), is the ``$\bigtriangleup$ structure''. And the yellow links connecting WA(L) to BA(L) and to v-preMA(L), respectively, is called the ``$\bigvee$ structure''. We use different colors for the link to distinguish their different frequencies ($f_{i}$) of activation (activate in $\#$ of subjects/the total $\#$ of subjects). This information is provided as in Table \ref{table:common1}: the ``$\bigtriangleup$ structure'' is found activated in all studied subjects. The ``$\bigvee$ structure'' is found in all the subjects in the bilinguals speaking Spanish group; this is different from monolingual group and bilinguals speaking English group, which only activated 75$\%$ and 50$\%$ of the time respectively.} 
	\label{fig:common1}
\end{figure}

The resulting common networks were constructed as described in Sec. \ref{sec:common}. In each group, the shared common network contained only nodes and links that are present across the majority of subjects. A visualization of the shared common networks at the fROI level are shown in Fig. \ref{fig:common1}. All groups' common networks have a similar fully connected structure involving pre-SMA, BA(L), and v-preMA(L) (we call it the ``triangle structure'', notated by $\bigtriangleup$) across all studied scans (n=8/8 in each group, 100$\%$). This structure is identified by looking at consistent edges connecting consistent fROIs across individuals. Therefore, this triangle captures the part of the individual functional network which goes beyond inter-subject variability that part of the individual network that is common across individuals. This structure is consistent with that found in our recent study of healthy individuals performing a different clinical pre-operative language task where we named it the ``core'' of the language network\cite{li2020core}.

Another structure involving WA(L)--v-preMA(L) and WA(L)--BA(L) (we called it the `` V structure'', notated as $\bigvee$) is present in 6/8 (75$\%$) subjects in the monolingual group, 4/8 (50$\%$) subjects in the bilingual speaking English group and 8/8 (100\%) subjects in the bilingual speaking Spanish group. This information is also summarized in Table \ref{table:common1}. These four regions are functionally connected with each other, with detailed modular link weights shown in Supplementary Tables \ref{table:common-M} - \ref{table:common-BS}.

\begin{figure}[!htbp]
	\includegraphics[width=0.45\textwidth]{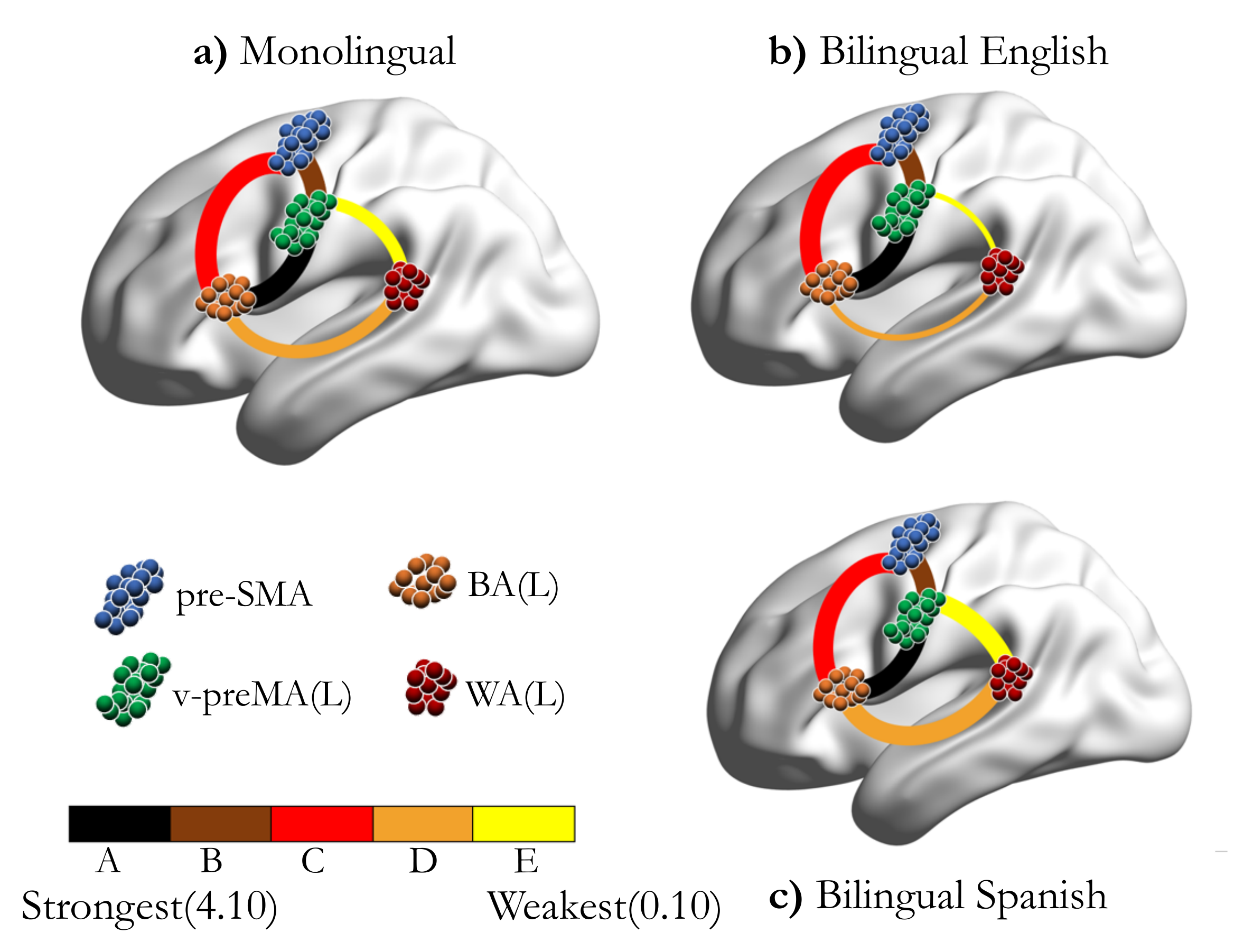}
	\caption{{\bf Common network}. Visualization of the shared common network across subjects in \textbf{a)} monolingual group, \textbf{b)} bilingual group speaking English and in \textbf{c)} bilingual group speaking Spanish, constructed by the methods described in Sec. \ref{sec:common}. Here, we show the sagittal view of the left brain. The modules are colored differently, and the color legend is provided right below panel \textbf{a)}. The link colors represent the $W_{ij}^{\rm C}$'s hierarchy strengths, within each group. The link colorbar is provided below the color legends of the modules. From Left to Right, we show the strongest (the largest $W_{ij}^{\rm C}$) to the weakest (the smallest $W_{ij}^{\rm C}$). The links between each fROI pairs are abbreviated as A-E, see Supplementary Tables \ref{table:common-M} - \ref{table:common-BS} for more details. The thickness of the links represents how frequently (in how many subjects) they appear. WA connects with preMA(L) and BA(L) in 75\% of monolingual subjects, 50\% of bilingual speaking English subjects and 100\% of bilingual speaking Spanish subjects, respectively.} 
	\label{fig:common}
\end{figure}

Though sample standard deviation in the link weights for each group is large relative to the mean, it is nevertheless evident that the common modular link weights are consistently larger for bilingual subjects speaking their native language, Spanish (L1) compared to when they speak the second language, English (L2), as shown in tables \ref{table:common-M} - \ref{table:common-BS}. A larger modular level link weight stems from the higher density in inter-fROI voxel level connections between common fROIs. The larger modular link  weight in the bilingual Spanish group coincides with a 100\% attachment level of the $\bigvee$ structure. The average common link weights of monolinguals tends to lie between that of the bilingual Spanish and bilingual English. The relative common modular link weights as determined by Eq. \ref{eq:common_weights} are indicated by the color bar in Fig. \ref{fig:common}. The thickness indicates the in group occurrence frequency $f_i$ of the link. 

\begin{table}[H]
	\caption[Frequency (in $\#$ of subjects/total $\#$ of subjects within one group) of each structure appear]{Frequency ($f_i$) of each structure appear}
	\label{table:common1}
	\centering
\begin{tabular}{|l|c|c|c|}
	\hline
	\multirow{2}{*}{} & Monolingual subjects & \multicolumn{2}{c|}{Bilingual subjects} \\ \cline{2-4} 
	& English task & English task & Spanish task \\ \hline
	\multicolumn{1}{|c|}{$\bigtriangleup$ structure} & 8/8 ($100\%$) & 8/8($100\%$) & 8/8($100\%$) \\ \hline
	\multicolumn{1}{|c|}{$\bigvee$ structure} & 6/8 ($75\%$) & 4/8($50\%$) & 8/8($100\%$) \\ \hline
\end{tabular}
\end{table}

 The shared common networks reveal a hierarchical ordering of link weight as shown in Fig. \ref{fig:common}. We denote the link BA(L)- v-preMA(L) by A, pre-SMA - v-preMA(L) by B, pre-SMA - BA(L) by C, BA(L) - WA(L) by D and WA(L) - v-preMA(L) by E. The link weight hierarchy ($A > B > C > D > E$) is consistent in all three groups. The results are consistent with our findings in our previous investigation conducted with 20 right-handed healthy controls \cite{li2020core}. In the previous study \cite{li2020core}, we named the four fROIs the language ``core'' sub-structure for the specific language task under study.
 
Though we have grouped the bilingual English and bilingual Spanish common networks separately, they are constructed from the same group of 8 subjects but performing the task in different languages. The monolingual group represents a different set of individuals. Therefore we expect an extra factor of inter-subject variability when making monolingual to bilingual group level comparisons as opposed to when making L1 to L2 based comparisons.

Note that not all pairs of core modules were directly connected. This is to be expected since, for example, pre-SMA and WA(L) have no known structural connections between them,  whereas the WA and BA are known to be connected by the Arcuate fasciculus \cite{carlson2012physiology}. The absence of a link does not convey direct information on the underlying structural connectivity due to intra and inter-subject variability in subject response to the task paradigm as noted above.
  
Our main findings are the differential attachment of the $\bigvee$-structure between groups and the higher common link weights in the L2 group compared to L1 in the $\bigtriangleup$-structure. The clinical implications of the observed differing structural properties of the FLNs in bilinguals as opposed to monolinguals with respect to the preoperative clinical task are discussed in Sec. \ref{sec:clin_rel}. 

\subsection{$K$-core analysis}\label{sec:kcore_result}

\begin{figure}[H]
	\includegraphics[width=0.48\textwidth]{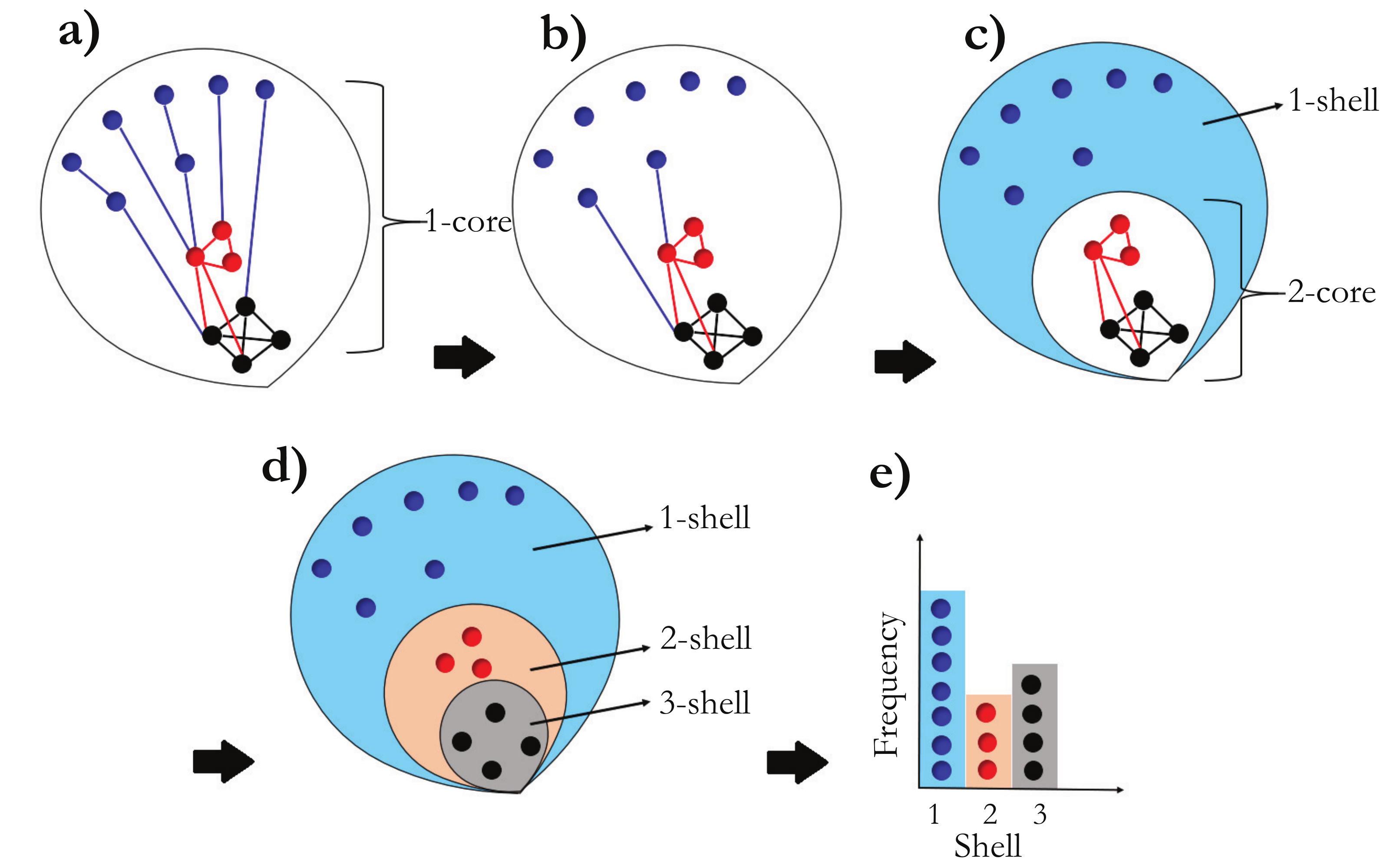}

	\caption{{\bf Schematic representation of a network going through $k$-shell decomposition}. At Step \textbf{a)}, it starts from 1-core (or $k$ = 1), which is also the whole network. At Step \textbf{b)}, we have disconnected all the nodes with a degree equal to one. After updating the degrees, now there are another two nodes with a degree equal to 1. We continue to disconnect those. At Step \textbf{c)}, we have disconnected all the nodes with a degree equal to 1 and there are no more nodes with a degree less than 2. The remaining graph (the nodes are still connected) make up the 2-core (or $k$ = 2). Those nodes which were removed from both \textbf{a)} and \textbf{b)} composed the 1-shell. The 1-shell and 2-core are exclusive from each other. We continue this process, we skip showing $k$ = 2 and 3, until Step \textbf{d)}, where all the nodes are disconnected; here, $k$ reaches its maximum, which is 3. Step \textbf{e)} is to plot $k$-shell histogram, where the horizontal axis is the shell number and vertical axis is the counts in each shell.}
	\label{fig:kcore_demo}
	
\end{figure}

The $k$-core of a given architecture is defined as the
maximal sub-graph, not necessarily one that is globally connected,
of all nodes having a degree (number of connections) of at least $k$. To partition the whole network into hierarchically ordered sub-networks, we perform a process to iteratively pruning all the nodes with degree $k$ until further removing is no longer possible (where the removing has caused the whole network collapse completely) \cite{burleson2020k}. The removed nodes are in the $k$-shell and the remaining subnetwork is called $(k+1)$-core. This final step will lead to finding the nodes in the maximum shell and the most connected sub-graph (maximum core) just before the whole network collapse. This process is called $k$-shell decomposition. We provide a brief explanation of $k$-core, the $k$-shell decomposition algorithm, and the meaning of the $k_{max}$ and $k$-shell occupancy histograms through a schematic $k$-shell decomposition process in Fig. \ref{fig:kcore_demo}. Starting from the whole network (1-core), as in \textbf{a)}, we begin with disconnecting all the nodes with degree equal to 1 ($k$ = 1) and then recalculate the degrees for each node left in the network and continue in removing nodes with updated degree of 1, as shown in \textbf{b)}. The disconnected nodes during this step is called nodes in 1-shell ($k$-shell, where $k$ = 1) and the remaining graph makes up 2-core ($(k+1)$-core, where $k$ = 1) as in \textbf{c)}. Then, we increase $k$ to 2 and repeat this pruning process until the whole network is dissembled. At this time, the final removed nodes are in the 3-shell ($k_{max}$-shell) and the final sub-network just before collapse is called the 3-core($k_{max}$-core).  

It has been shown that for networks with positive couplings (positive link weights), the $k$ max core is the component of the system that is most resilient with respect to network failures, where in this case a failure means a reduction of the link weight (potentially due to brain tumor invasion or physical resection) \cite{morone2019k}. It turned out that all the thresholded voxel-voxel BOLD time series correlations defining the link weight for our experimental task paradigm were positive. Thus, conducting $k$-core analysis would reveal the most robust component of the functional language architecture in the healthy monolingual and bilingual subjects. 

\begin{figure}[htpb!]
	\includegraphics[width=0.45\textwidth]{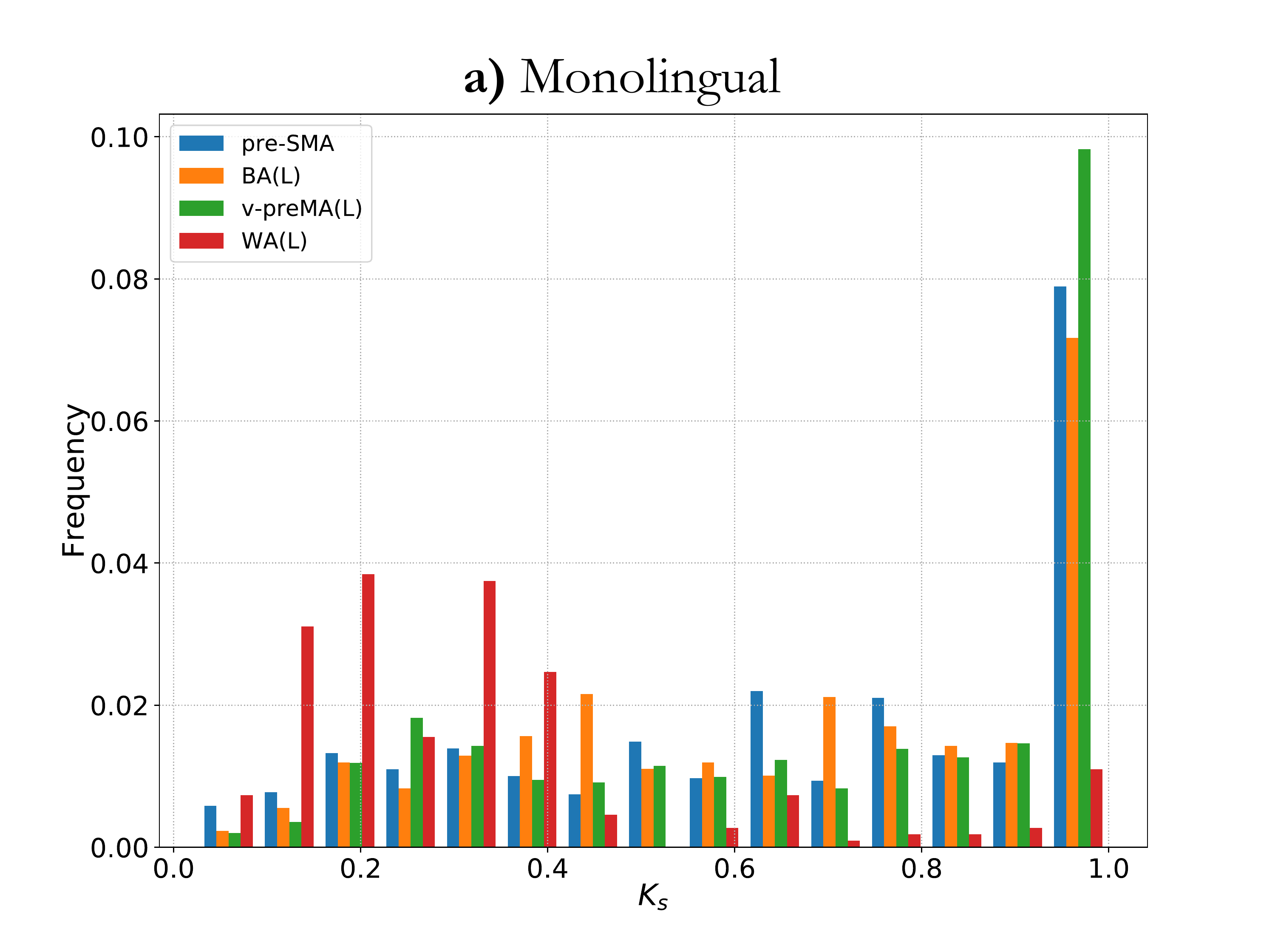}
	\includegraphics[width=0.45\textwidth]{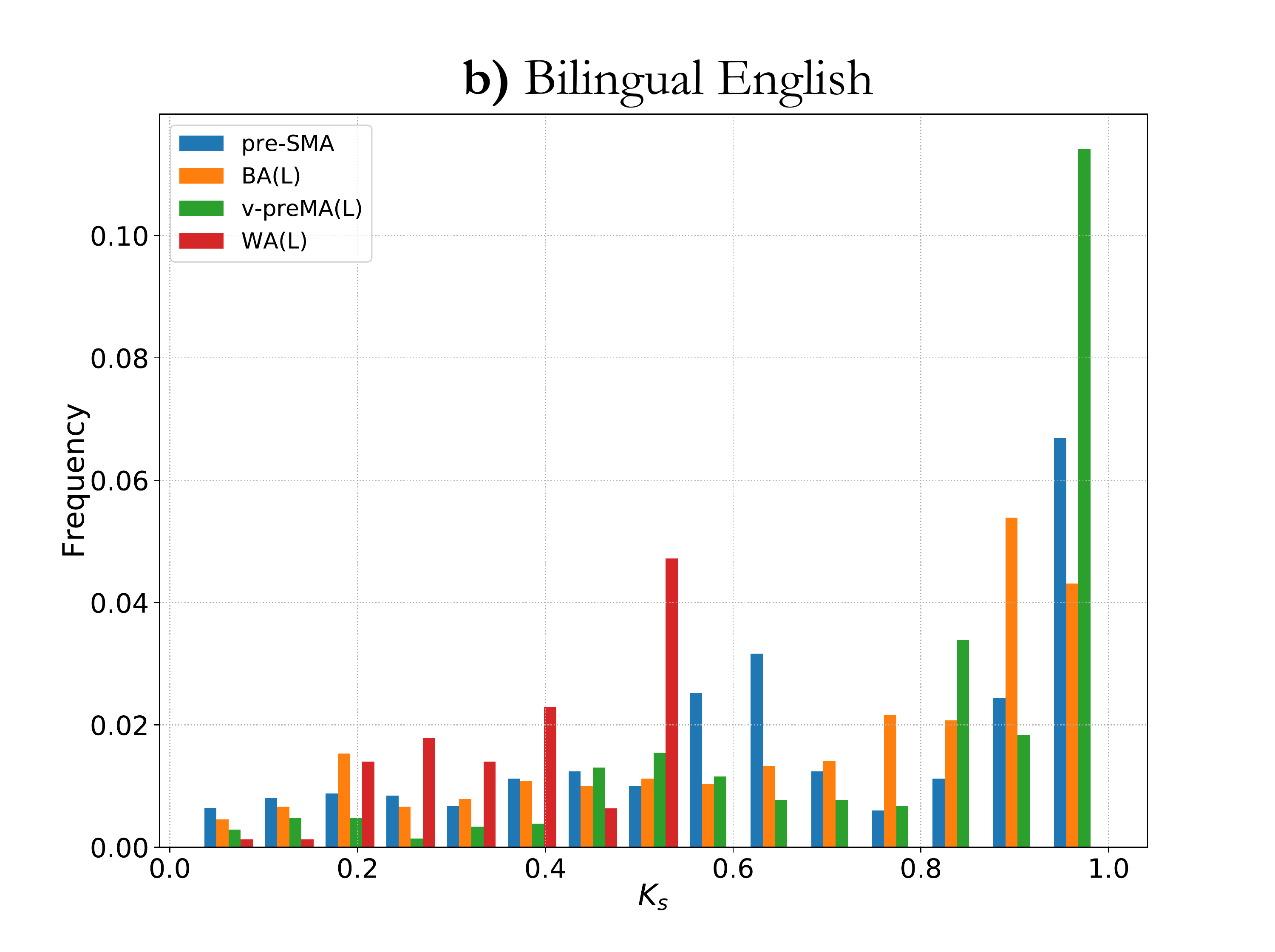}
	\includegraphics[width=0.45\textwidth]{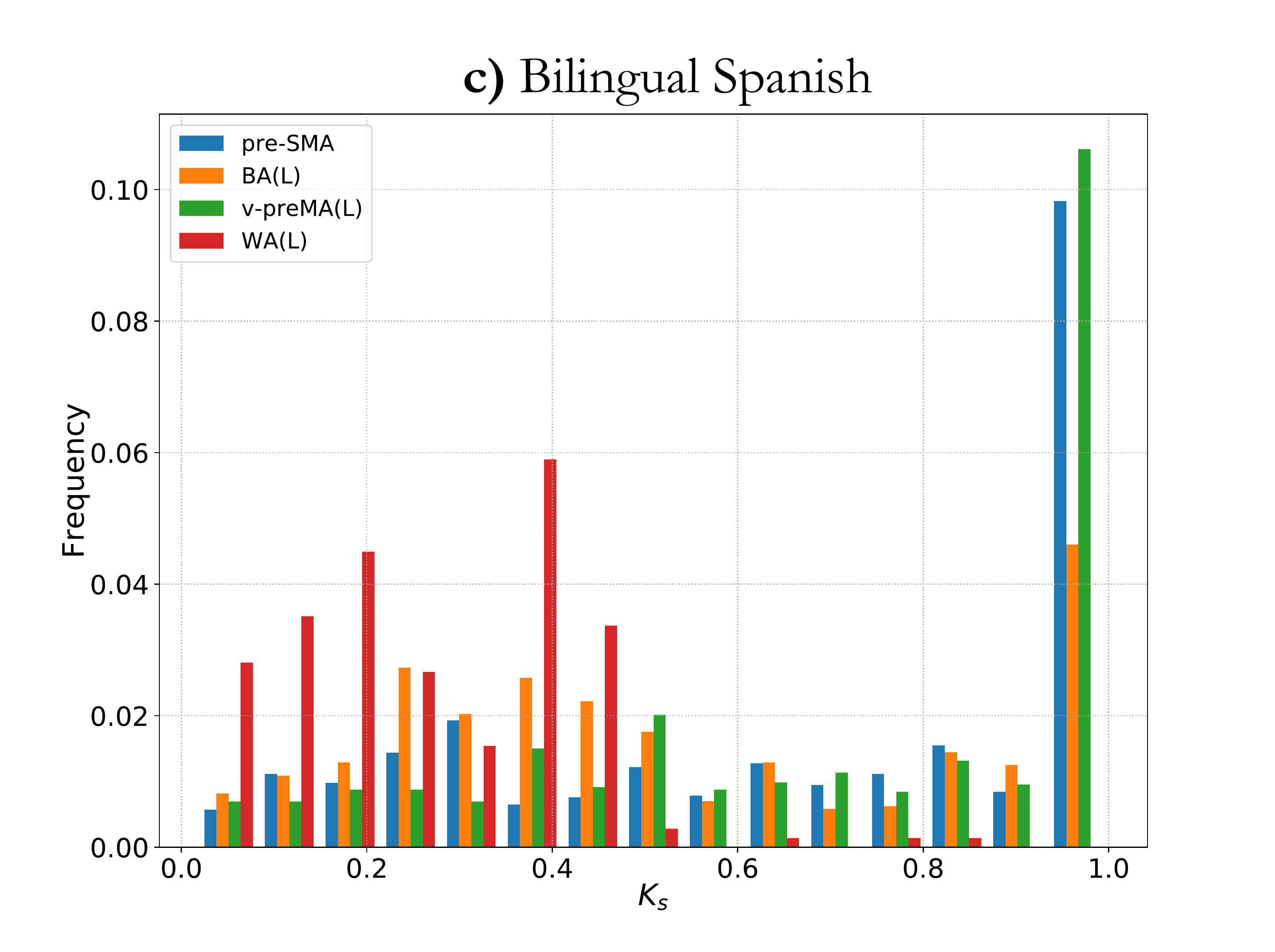}
	\caption{{\bf $k$-shell occupancy for all three groups.} Different colors represent nodes belonging to different modules and the color legend is shown at the upper left of the figure. In \textbf{a)} - \textbf{c)} pre-SMA, BA(L), and v-preMA(L) peak at the maximum shell; however, WA(L) occupies mostly in middle or low shell. Especially in \textbf{b)}, WA(L) does not occupy the higher ($k_{s}$ > 0.5) shell at all.}
	\label{fig:kshell}	
\end{figure}

The $k$ shell (occupancy) histogram provides important and direct insights to the network structures. Therefore, for each group, we calculate the $k_s$ shell occupancy for each node in the individual network \textit{at the voxel level}. Then, we normalize $k_{s}$ by the maximum shell number found in each individual network, $k_{max}$, so that $k_{s}$ ranges from 0 to 1, where $k_{s}$ = 1 is the maximum shell ($k$ max core). Then, we collect all the individual networks' nodes together, regardless of which subject they came from, and placed them into 15 bins according to their $k_{s}$ values. Then, we group the nodes in each bin by the modules they belonged to and finally plotted one unique $k$-shell occupancy histogram for each group. The histograms are shown in Fig. \ref{fig:kshell} for the four modules from the common shared core of the FLN: pre-SMA, v-preMA(L), BA(L), and WA(L) (colored differently). Panels \textbf{a)} - \textbf{c)} display the $k$ shell histogram for monolinguals, bilinguals speaking English, and bilinguals speaking Spanish, respectively.

\begin{figure}[htpb!]
	\includegraphics[width=0.5\textwidth]{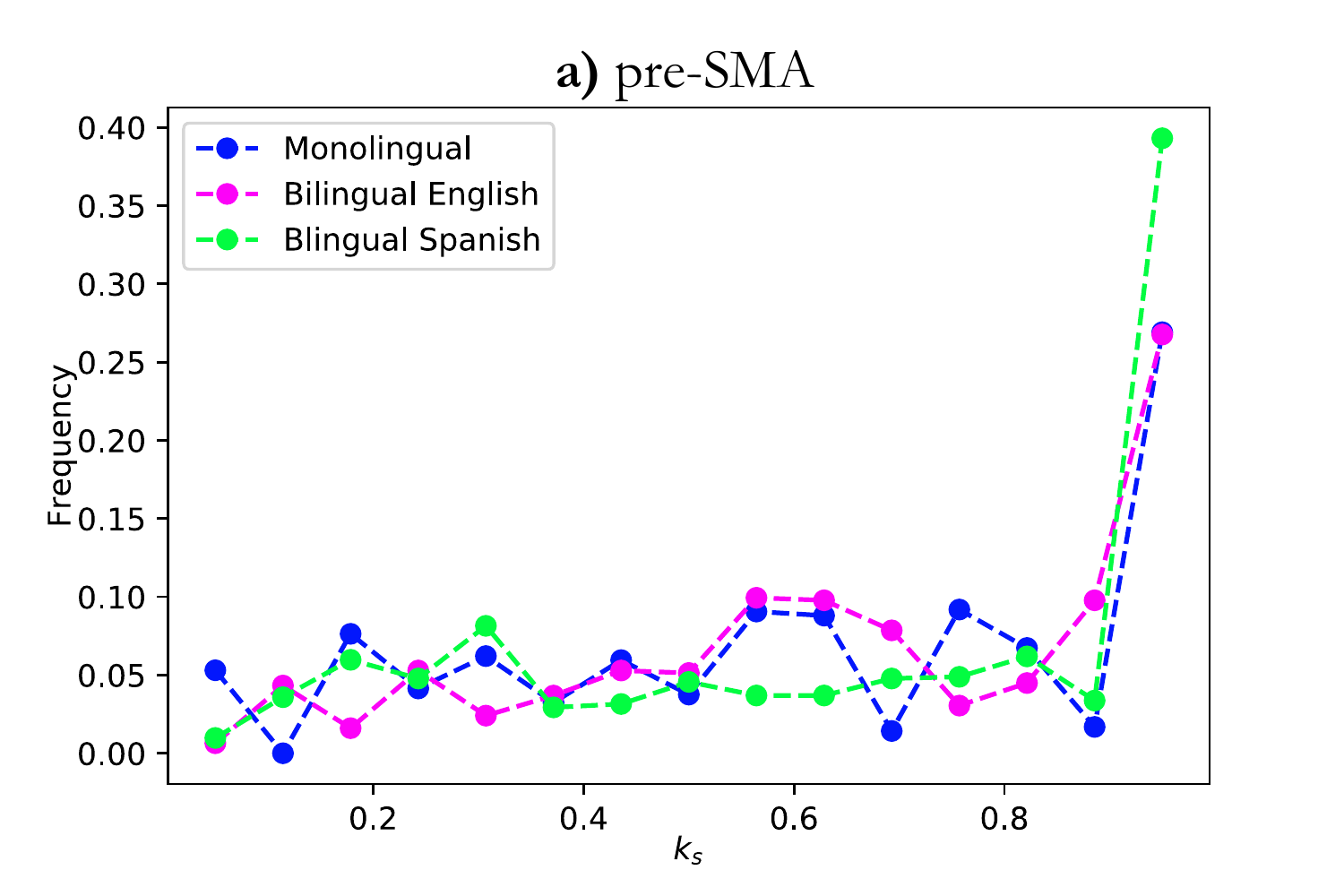}
	\includegraphics[width=0.5\textwidth]{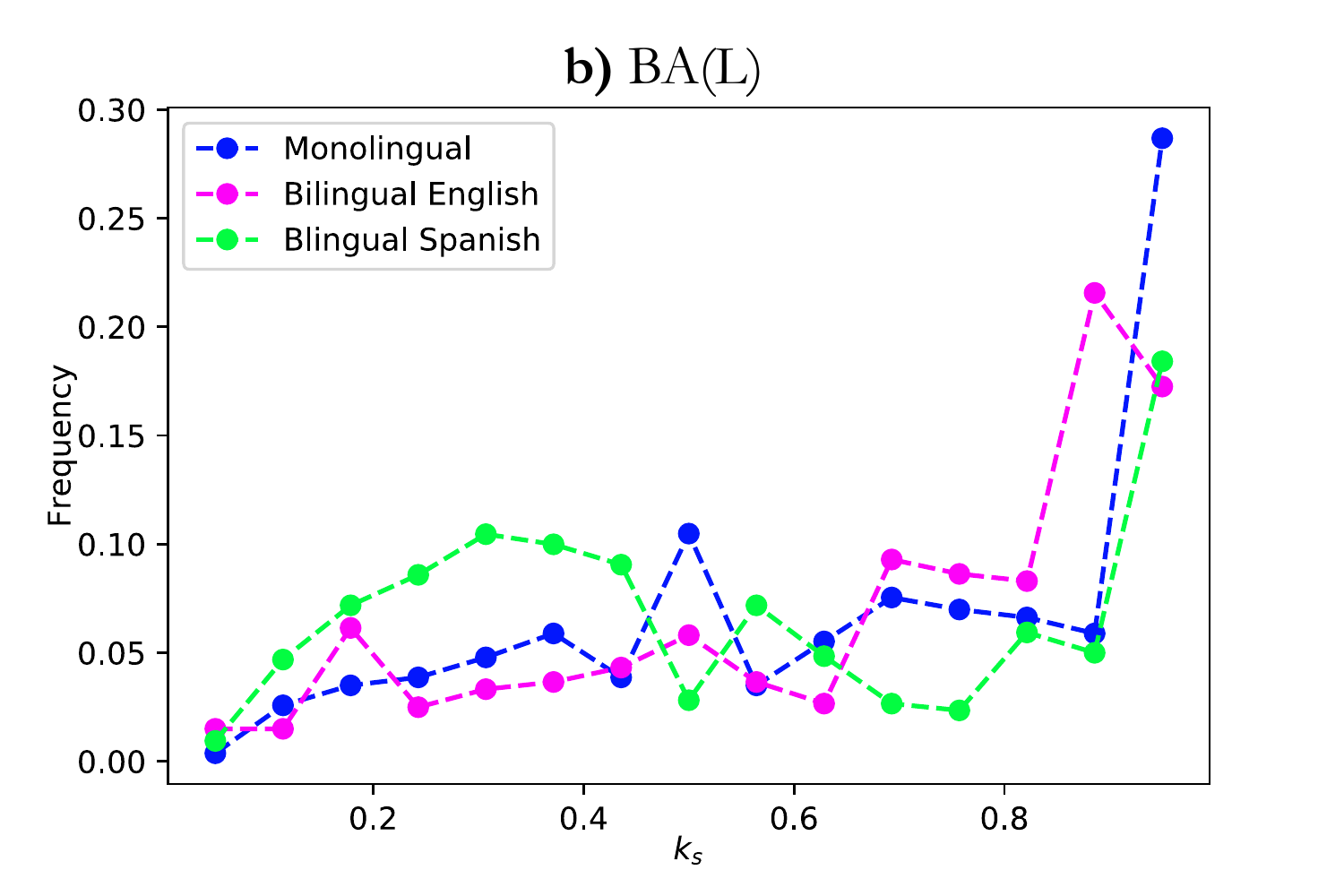}
	\includegraphics[width=0.5\textwidth]{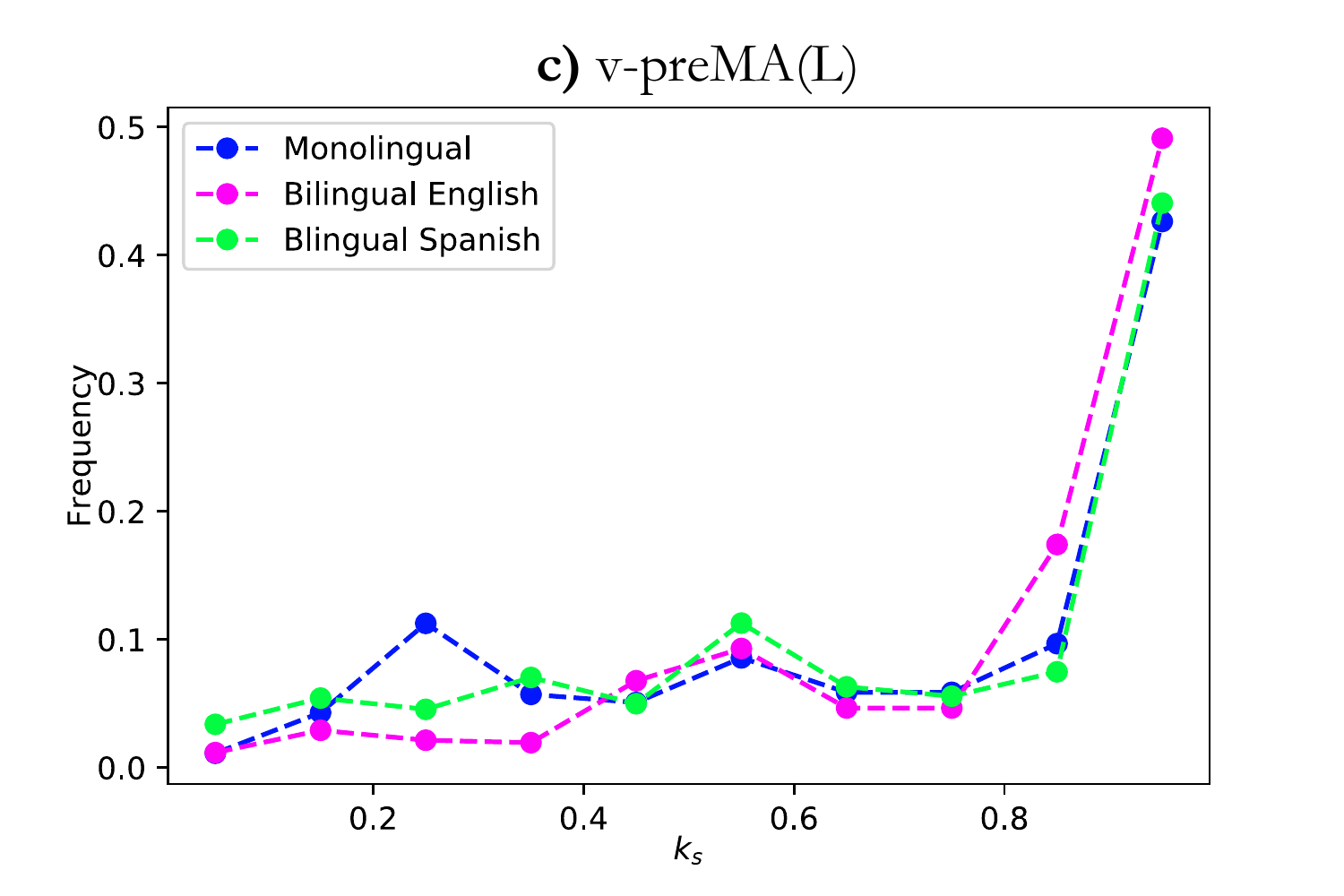}

	\includegraphics[width=0.5\textwidth]{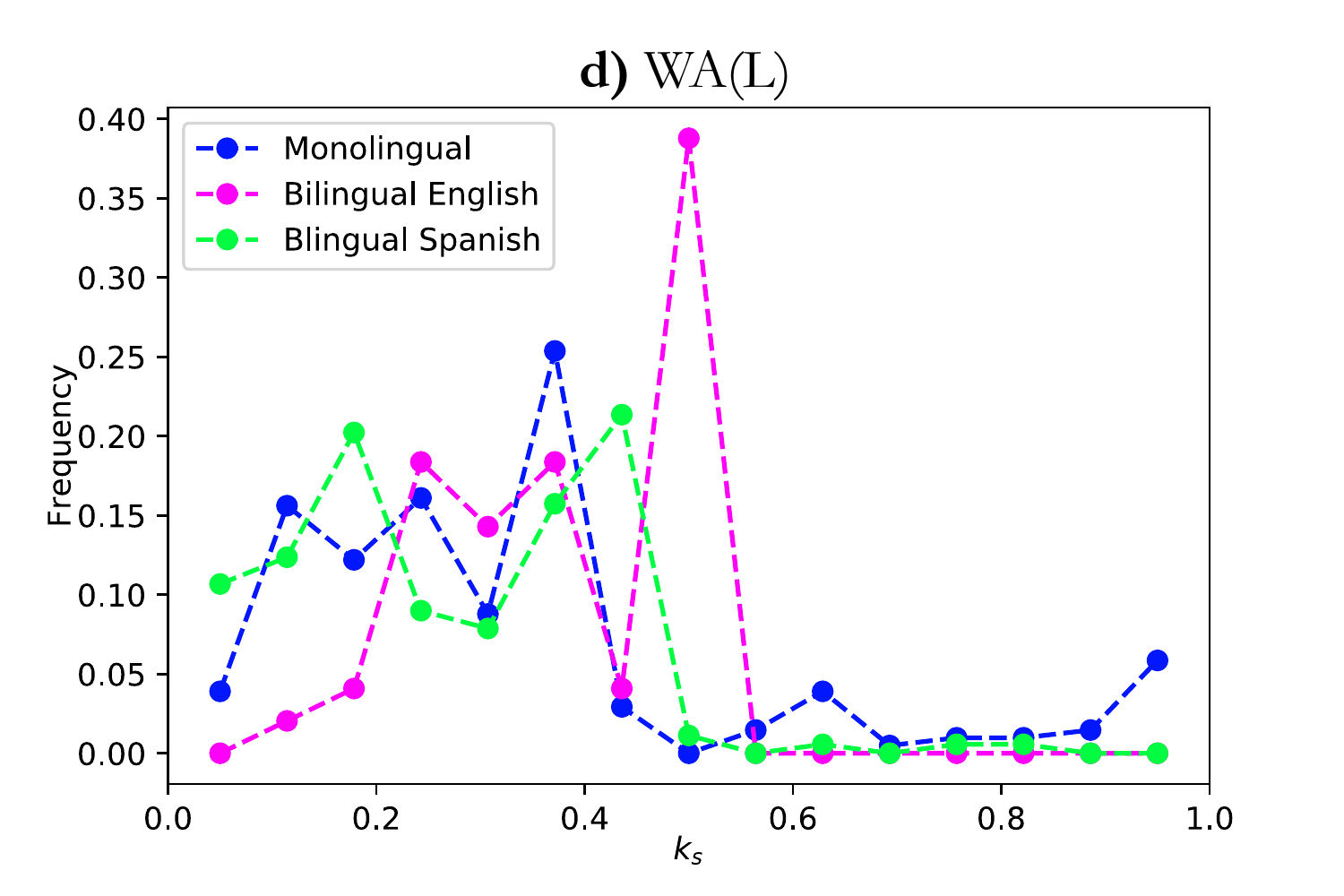}
	\vspace{-1cm}
	\caption{{\bf \textbf{a)} pre-SMA, \textbf{b)} BA(L), \textbf{c)} v-preMA and \textbf{d)} WA's $k$-shell occupancy for all three groups.} This time, different colors represent different groups. Notice that, the three colored curves are most distinctive in panel \textbf{d)}. For this panel only, the most distinctive behaved group is Bilingual English but all three groups appear to be quite different.}
	\label{fig:WA_distributions}	
\end{figure}

In parallel, we plot each module's $k$ shell histogram separately, as shown in Fig. \ref{fig:WA_distributions}. Additionally, in order to validate the results, we conduct a statistical analysis to determine whether the differences in $k$-shell occupancy distributions between groups are statistically significant. To this end, we compute the sum of squared errors (SSE or residuals) between each distribution pair for each of the four modules.

\begin{figure}[htpb!]
	\includegraphics[width=0.5\textwidth]{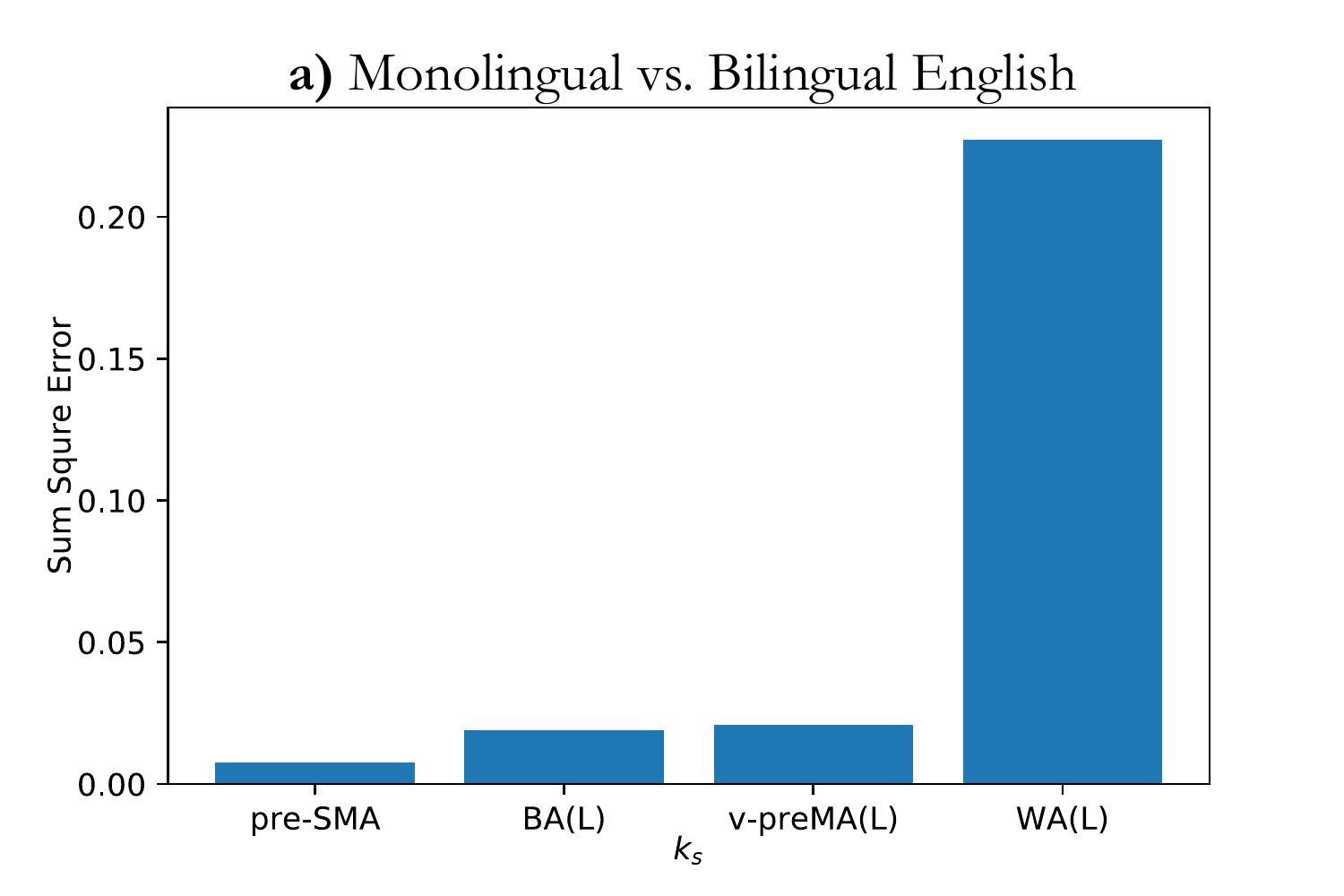}
	\includegraphics[width=0.5\textwidth]{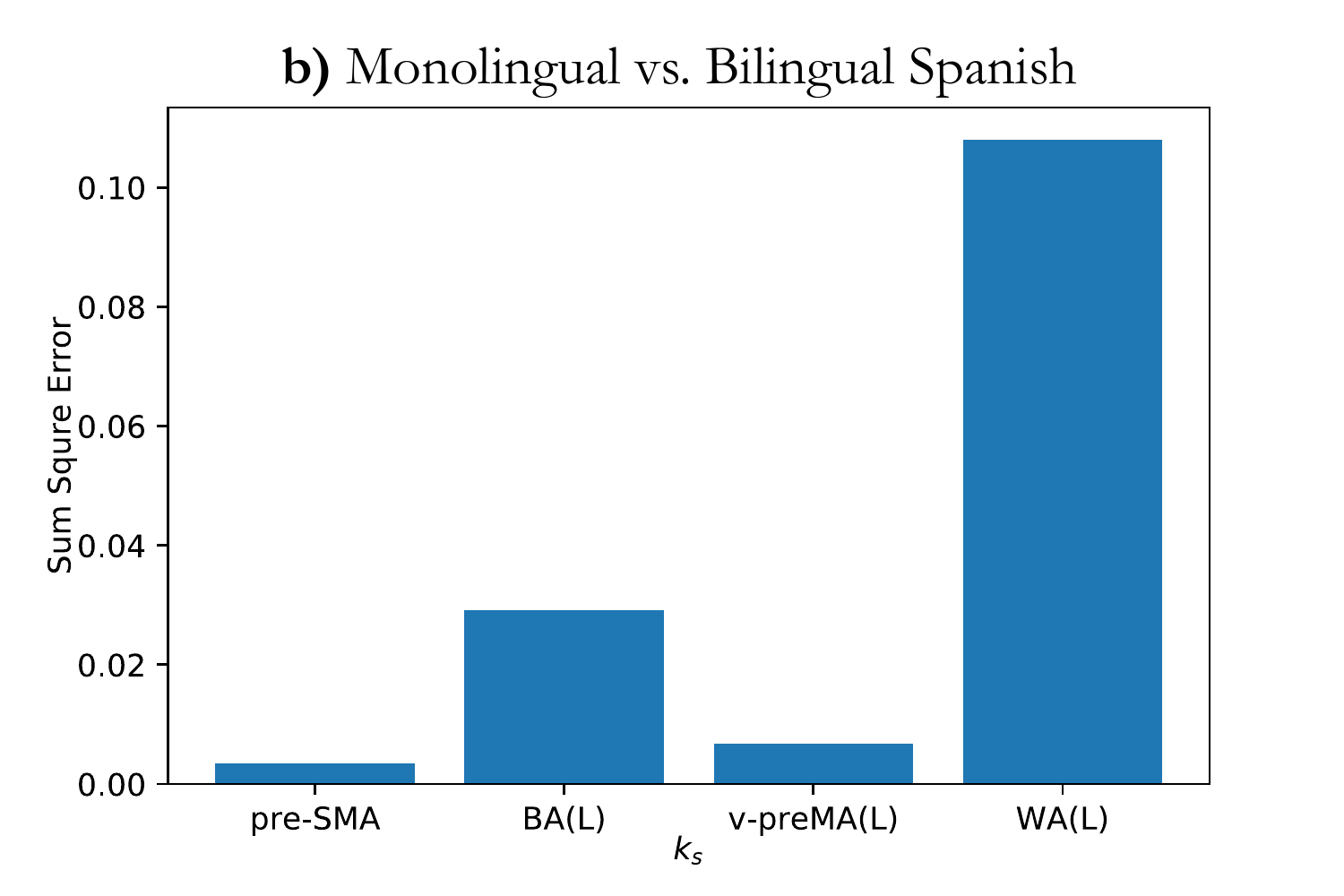}
	\includegraphics[width=0.5\textwidth]{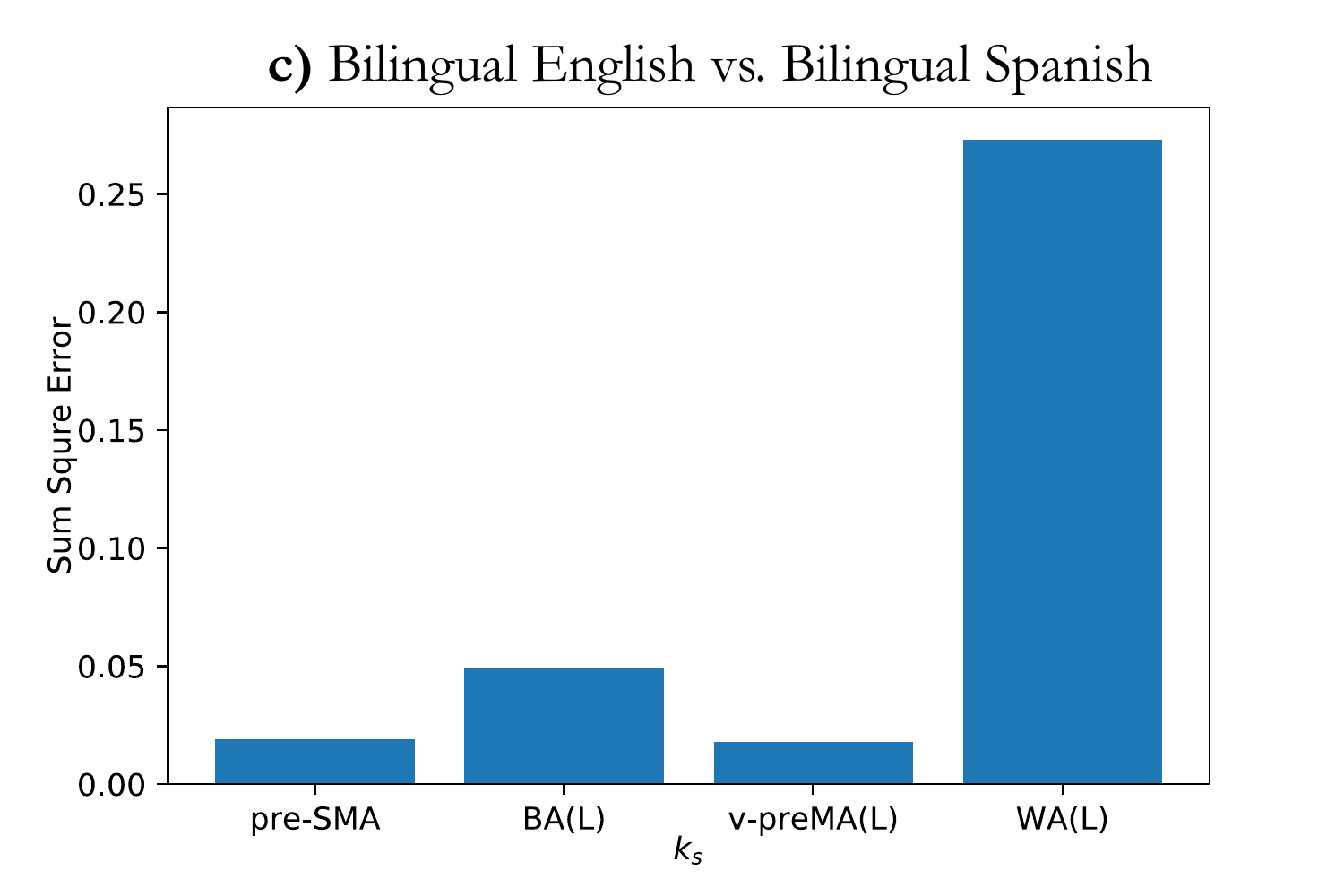}
	\caption{{\bf Sum of square errors (SSE) of $k$-shell occupancy for each module.} For all three pair of groups, as shown in panels \textbf{a)} to \textbf{c)} WA(L)'s distribution has the largest SSE (5-10 times greater) compared to the ``core'' modules. Bilingual English and Spanish's SSE in WA comparison, as seen in panel \textbf{c)}, has the largest value while monolingual and bilingual Spanish has the smallest SSE in WA.}
	\label{fig:squared_errors}	
\end{figure}

First, we observe that, the shell occupancy of the ``core'' modules - BA(L), v-preMA(L), and pre-SMA are quite similar in all three groups: most of the nodes in these three modules occupy the maximum shell ($k_{max}$). This can be seen in Fig. \ref{fig:kshell} \textbf{a)} - \textbf{c)} and Fig. \ref{fig:WA_distributions} \textbf{a)} - \textbf{c)}. This is also confirmed with statistical analysis in Fig. \ref{fig:squared_errors}. By similarity in distribution, we mean that the sum of squared errors is relatively small (no greater than 0.05) in the core modules compared to the WA(L) which displays much greater sums of square residuals as can be seen later. 

Secondly, in contrast with the ``core'' regions, the occupancy of WA(L) is fundamentally different since most of the nodes do not occupy the maximum shell. Rather, it is distributed among the smaller shells ($k_{s} <= 0.5$) in the peripheral of the network. This is observed consistently (in Fig. \ref{fig:kshell} \textbf{a)} - \textbf{c)} as well as Fig. \ref{fig:WA_distributions} \textbf{d)}) among all three groups: monolingual, bilingual speaking English and bilingual speaking Spanish. This is supported by the SSE analysis as in Fig. \ref{fig:squared_errors} \textbf{a)} - \textbf{c)} that shows that WA(L) is five to ten times greater in the SSE compared to the ``core'' modules.
It is worth to mention the above two points are in agreement with recent findings that WA(L) may belong to the lower shells \cite{li2020core} compared with the ``core'' regions.

Furthermore, comparing the groups for WA(L) occupancy, we notice some subtle differences between the three groups. For example, as shown in Fig. \ref{fig:WA_distributions} \textbf{d)}, the occupancy of the WA(L) in bilingual speaking English group presents a peak at $k_{s} = 0.5$ and does not extend to higher shells. This result suggests that the WA may be less resilient in its attachment to the FLN in the bilingual subjects speaking English - the subject's second language (L2). 
In addition, WA(L)'s occupancy in the bilingual speaking English group shows less occupancy in the peripheral outer shells (small $k_{s}$) than the other two groups. Visually, the occupancy distribution of WA(L) in monolingual and bilingual speaking Spanish appear more similar to each other than to the bilingual speaking English group. These apparent difference are corroborated by the SSE analysis shown in Fig. \ref{fig:squared_errors} \textbf{a)} - \textbf{c)}. 

To summarize the results from the SSE analysis, the ``core'' $k$-shell distributions tends to be similar between groups whereas for the WA, the $k$-shell histogram varies significantly between each group pair.
Besides, we find a larger SSE ($\sim$ 0.2-0.25) in the occupancy of WA when we compare the bilingual speaking English group to either the monolingual group or the bilingual speaking Spanish group, as shown in Fig. \ref{fig:squared_errors} \textbf{a)} and \textbf{c)}.
On the other hand, the SSE of WA drops to half of this value (SSE $\sim$ 0.1) when we compare the monolingual with the bilingual speaking Spanish group, implying that the occupancy of WA is more similar between the monolingual with the bilingual speaking Spanish group than with the bilingual speaking English group.
  
 We note that this SSE analysis is not a formal hypothesis test so this data provides evidence for statistical significance but is not an established method. However, what the residual analysis does suggest is that the observed occupation behavior of WA is less likely to be due to random noise in the data, and more likely reflects real underlying trends in the data.

\subsection{Other centrality measurements}\label{sec:centrality-result}

In this study we focus on the $k$-core centrality measure, however, there are several other centrality measures in complex networks. Each type of centrality measure offers a different perspective on the network structure and may provide additional important information.

Centrality measures relate to the influence of nodes with respect to functional integration and resilience in the event of network failure events such as the destruction of a node or link. Besides the $k$-core, we also computed the classical centrality measures in order to obtain additional characteristics of the functional network architectures. We measured three types of classical centralities: degree, closeness, and betweenness as summarized in Supplementary Table \ref{table:centrality}. Degree centrality simply measures the number of links attached to a node. Closeness centrality is inversely proportional to the average length of the shortest path between a node and all other nodes in the graph, while betweenness centrality measures the fraction of all shortest paths between other nodes passing through a given node. 

We ran algorithms to calculate each type of centrality of all individual voxel-scale networks. Then, we obtained an averaged value over nodes inside each of the common network's fROIs. The final value displayed in the text represents the average across all the subjects within a group. As shown in Supplementary Table \ref{table:centrality} the WA(L) displays the highest betweenness centrality, while its degree centrality and closeness centrality are weaker than the other three members of the ``core''.

\section{Discussion}\label{sec:discussion}

The goal of our study is to identify essential and non-essential language areas in primary and secondary languages to guide intra-operative procedures. This identification is supported with a network analysis to study the structure and interconnections between essential areas for both monolingual and bilingual groups. We conducted this study by first constructing FLNs on the individual level and then we aggregated these results by group to identify group structures and to highlight differences between groups. 


\subsection{Common networks}

We observe that there are both similarities and differences between monolinguals and bilinguals functional networks. All groups share a core network composed of a resilient ``triangular structure''. The ``triangular structure'' connects also to WA to form the ``V structure'', with different degrees at the group level: 8/8 bilingual subjects speaking Spanish, 6/8 monolingual subjects, and 4/8 bilingual subjects speaking English. Bilingual English subjects display the smallest common link weights while the same subjects performing the task in Spanish have the largest common link weights. These results reflect the higher clinical task engagement of language processing systems related to L1 when compared to L2.

The hierarchy of strengths between the three clusters, $A > B > C > D > E$, is consistent across all three groups, as shown in Fig. \ref{fig:common} and Supplementary Table \ref{table:common-M} - \ref{table:common-BS}. This is also consistent with our previous analysis performed in 20 right-handed healthy subjects \cite{li2020core}. This consistent hierarchy may predict the amount of information traffic flowing between each interacting modules. The low connectivity weight between ventral preMA and WA (see Supplementary Tables \ref{table:common-M}-\ref{table:common-BS}) may be explained by the increased distance between the two structures. 

The areas wired in the core may be pivotal in language
production from the point of view of hodotopy, where the function of single areas is integrated in a network perspective \cite{duffau2014re}.
Tate et al. \cite{tate2014probabilistic} demonstrated that speech arrest localizes more often to the ventral preMA(L) instead of the classical BA(L) during intraoperative direct cortical stimulation. Also, tumoral invasion generates more speech deficits by infiltrating the left ventral preMA than BA.

This area is considered to serve articulatory planning in speech production, consequently leading of negative motor response during stimulation \cite{rech2019probabilistic}. Premotor activations are also connected to cognitive processes \cite{rizzolatti2002motor,cerri2015mirror}, and the ventral premotor belongs to the action-observation circuit as human analogue of F5 in macaque monkeys \cite{pattamadilok2016contribution}. Our results confirm the evidence for the important role played by the left ventral preMA in language processing, regardless of which language people speak. The premotor cortex is known to participate in bilingualism under the name of DLPFC. Nevertheless, its participation in the core and its relationships with other components are not well described. As mentioned in Sec. \ref{sec:introduction}, the premotor should have a function to control language selection. In our results, the consistent activation of preMA can be partly explained by its role in the extended BA, as discussed in our previous work \cite{li2020core}, which may be valid regardless of the language spoken. Particularly, the strong link between opercular BA and ventral premotor cortex should constitute a hub for language production by connecting with SMA \cite{corrivetti2019dissociating}.

Although the correspondence between structural and functional connectivity is not fully understood yet \cite{bullmore2009complex, honey2009predicting}, our results may be supported by structural evidence. This refers to known white matter bundles connecting the network core nodes, such as the frontal aslant tract and the dual pathways of language \cite{li2020core}.

As a minor difference between the groups, we find more frequent activation of secondary language areas in bilinguals. This result aligns with previous studies that show that the left Caudate and Angular Gyrus is relatively more involved with bilingualism \cite{seo2018bilingual,li2016functional}.

\subsection{$K$-core} \label{sec:kcore_discussion}

With respect to the $k$-core analysis, we see in Fig. \ref{fig:kshell} that the bilingual-english group WA does not populate the lower $k$ shells and that when comparing monolinguals versus the bilingual spanish, WA populates more in the lower $k$ shells of the bilingual spanish group. Monolinguals populate the $k$-max core whereas the bilingual groups do not. WA populates the $k$-max core less than in the core but still significantly.  

The $k$-core max needs to be highly connected to other highly connected nodes, whereas the the low shells are similar to dangling ends of the system. In general, the $k$ shell represents a hierarchy of nodes. Thus we can say that bilingualism is manifests itself in a reduction of the most important nodes, at least as far as WA is concerned. These results cannot be obtained with methods based only on the activity of the brain, and one needs to do the network analysis to differentiate the three states of the WA. That is, our network analysis finds a difference in the role of the WA while using only activity will not.

We consistently observe that the greatest portion of the three modules pre-SMA, BA(L), and v-preMA(L) occupied the largest, $k$-max shell. This suggests that the triangle's modules are the most resilient part of the network, which prevents cascading failures in the event of network failures \cite{rubinov2010complex,dorogovtsev2006k,morone2019k}. This sub-structure may thus prevent network collapse in the event of removal of links as caused by pathological conditions and/or subsequent surgical intervention. In either case, damage to these core links may result in damage to the language network in an irreversible way.

It should be noted that the presence of WA in the weaker $k$ cores does not necessarily imply that WA is less important to the language network. Rather, weakly connected nodes can sometimes form pivotal roles in the network processes. Morone \textit{et al.} \cite{morone2015influence} have shown that the nodes with low degrees are sometimes the most important essential nodes when they hold the keys to connections between hubs (modules). In this context, the $k$-core results might indicate that WA(L) has at least a distinctive functional nature to the other three core members of the FLN, with respect to the network path structure. Recent studies evidenced distinct anatomical substrates for the motor-speech and lexico-semantic systems, speculating the existence of a double triangular network serving lexico-semantic processes and speech articulation \cite{corrivetti2019dissociating}. The areas wired in the network show an intriguing correspondence with our results and structural evidence from the literature \cite{catani2012short}.

The results from our $k$-core analysis may also be supported from structural evidence. We found a significant difference in the $k$ shell distribution of WA(L) nodes when performing the L1 task compard to L2. As well, the other common fROIs occupied the maximum shell with different proportions in L1 versus L2 as shown in Fig. \ref{fig:kshell}. In fact, there is evidence for significant differences in structural connections between monolinguals and bilinguals. Diffusion Tensor Imaging (DTI) studies have demonstrated that the representation of uncinate fasciculus, which connects the deep opercular cortex with the superior anterior temporal lobe \cite{friederici2011brain} is increased in bilinguals \cite{wong2016neurolinguistics,luk2011lifelong}. Similarly, increased fractional anisotropy of the superior longitudinal fascicle, connecting the preMA(L) with the superior temporal gyrus, has been reported in bilinguals \cite{wong2016neurolinguistics,luk2011lifelong}.

\subsection{Other centrality measurements}
As shown in Supplementary Table \ref{table:centrality}, the WA(L)'s betweenness centrality scores are higher than those of the other three core members, indicating that information tends to often pass through the more centrally located WA(L) in the functional language network. This is despite the fact that nodes in the WA(L) have lower degree centrality (fewer average connections) as well as lower closeness centrality (longer path length). The longer path length of the WA(L) nodes are compatible with the larger anatomical separation between WA(L) and the other common core fROIs. The fewer average connections of WA(L) nodes is compatible with the $k$-core results for the WA(L). Yet, WA(L)'s betweenness score shows that it acts as bridge between other more highly conected components of the language network. This finding illustrates the complex architecture of the functional language network. Generally, real networks cannot be simply characterized by a single network theoretical measure but will tend to display several properties, for example, an interplay of small-world and scale-free features \cite{gallos2012small,rozenfeld2010small}.

\subsection{Clinical relevance}\label{sec:clin_rel}

When the clinical task is performed by bilingual subjects using L1, the common inter-modular connection weights are stronger and the frequency of certain links, notably in the $\bigvee$-structure, increase as shown in Table \ref{table:common1}. This indicates that a more robust pre-operative language map may be obtained by conducting the fMRI language task in the subjects native language, L1, rather than L2. These data may also help to optimize pre-surgical planning on an individual basis.

In the clinical setting, information about the stability of pivotal nodes obtained from the $k$ core analysis may play a role in guiding the neurosurgeon to avoid interrupting essential functional links. In fact, the resection of primary language areas does not necessarily cause the expected spectrum of language deficits. This has been reported also for WA \cite{petrovich2004isolated,sarubbo2012resection} and may be due to the involvement of subcortical paths and compensatory mechanisms \cite{fisicaro2016cortical}. If confirmed by further investigation, different engagements of a node by the language core may correlate with different surgical outcomes across different populations.

The evidence of a network core for language shared by monolinguals and bilinguals may provide useful information in pre-surgical planning: as shown by our $k$-core evaluation, disconnection of the core nodes by cortical resection or interruption of subcortical pathways may lead to the collapse of the language function. 

The common network of bilingual subjects speaking English shows limited participation of WA. Such evidence may support the introduction of language tasks optimized for bilinguals in clinical practice, even for subjects fluent in English. The variable participation of WA to the core may be related to the clinical task being proctored in a sub-optimal language, L2, for bilinguals when it should ideally be proctored in L1.

While the common functional network and its $k$ core structure serve to illustrate typical patterns in healthy patients it is important to remember that for clinical purposes each individuals language map is unique and may not conform to these patterns. Damage to brain tissue from disease may disrupt the core network and potentially shift the position of core fROIs \cite{litranslocate2019}. 

Our evidence may suggest that clinicians should consider employing the pre-operative language task in either L1 or both L1 and L2 to ensure that a robust and accurate language map is obtained. This would certainly be preferable to the current clinical standard of employing the task solely in English. Obviously, practical considerations apply, and the task must be redeveloped in several languages corresponding to the dominant ethnic populations in the regions surrounding each clinic.

\subsection{Limitations}

The main limitation of this study is the sample size. This results in difficulty to ascertain the statistical significance of the common network properties due to inter and intra subject variability in response to the clinical task. Statistical uncertainty due to small sample size also propagates into the $k$-core measurements since we focused on the core fROIs in our analysis. We suggest that these results be probed with better statistical significance in future studies with larger sample size. Additionally, we employed only one language task in our study. Clearly, a single language task cannot successfully and completely model the entirety of the language network. 

\section{Summary}

Though many studies have investigated bilingualism through fMRI, this is the first study to consider the effects of bilingualism when subjects perform pre-surgical language tasks. We sought to establish a healthy benchmark for the FLN in both monolinguals and bilinguals to assist clinical decisions, with the aim of providing insights to clinicians about the diagnosis and treatment for ethnic or minority groups who speak a second language other than English. 

We discussed the main results of our study, that both monolingual and bilingual subjects share a common language network formed by BA, preMA, and pre-SMA, which occupy the $k_{max}$ shell and show features of a central core for language across groups, consistent with our previous results on healthy subjects \cite{li2020core}. Moreover, WA is engaged by the network core with variable extent across groups (8/8 bilingual subjects speaking Spanish, 6/8 monolingual subjects, and 4/8 bilingual subjects speaking English), reflecting different $k$-core occupancies, with the major difference being that the bilingual L2 groups' nodes occupy only the lower half of $k$ shells. The bilingual L2 group is also revealed to show weaker connection strengths between core fROIs compared to when they perform the task in L1. 

These results may influence fMRI task choice and interpretation, with possible impact on therapeutic planning. In order to optimize the pre-operative language map for bilingual patients, we suggest that clinicians consider implementing the language task in L1 instead of L2. This recommendation is based on the higher clinical task engagement of language processing systems related to L1 compared to L2 as evidenced by the higher common network link weights and more frequent involvement of the ``V-structure" as discussed above.

\section{ACKNOWLEDGMENTS}

The funding support for this study was provided by
the National Institute of Health (NIH), NIH-NIBIB R01
EB022720-01 (Makse and Holodny, PI's), U54CA137788,
U54CA132378, P30CA008748, the National Science
Foundation (NSF) NSF-IIS 1515022 (Makse, PI), ISSNAF
imaging chapter award 2018 (Pasquini, PI) and
ESOR Bracco clinical fellowship 2018 (Pasquini, PI).
\section{Data availability}
This data set is publicly available at:
\url{http://www-levich.engr.ccny.cuny.edu/webpage/hmakse/brain/}.

\bibliographystyle{myunsrt}
\bibliography{HEALTHY}

\setcounter{section}{0}
\setcounter{figure}{0}

\renewcommand{\figurename}{{\bf Supplementary Figure}}
\renewcommand{\tablename}{{\bf Supplementary Table}}

\setcounter{table}{0}
\setcounter{figure}{0}
\renewcommand{\thetable}{S\arabic{table}}
\renewcommand\thefigure{S\arabic{figure}}
\renewcommand{\theHtable}{Supplement.\thetable}
\renewcommand{\theHfigure}{Supplement.\thefigure}

\clearpage
\onecolumngrid

\section*{\bf {\Large Supplementary Information}}

\section{Active areas}
\begin{table}[H]
	\caption[Active areas across subjects]{Active areas across subjects.}
	\label{table:activeareas}
	\centering
	\begin{tabular}{|c|c|c|c|c|}
		\hline
		Activated areas                  & Abbreviations & \begin{tabular}[c]{@{}c@{}}Activated in \\ \# of subjects\\ in monolingual\end{tabular} & \begin{tabular}[c]{@{}c@{}}Activated in\\ \# of subjects\\ in bilingual English\end{tabular} & \begin{tabular}[c]{@{}c@{}}Activated in \\ \# of subjects\\ in bilingual Spanish\end{tabular} \\ \hline
		Angular Gyrus(L)                 & AngG(L)       & 12.5\%                                                                                       & 50\%                                                                                             & 37.5\%                                                                                              \\ \hline
		Broca's Area(L)                  & BA(L)         & 100\%                                                                                        & 100\%                                                                                             & 100\%                                                                                              \\ \hline
		Broca's Area(R)                  & BA(R)         & 25\%                                                                                        & 0\%                                                                                            & 0\%                                                                                             \\ \hline
		Caudate(L)                       & Caudate(L)    & 12.5\%                                                                                       & 12.5\%                                                                                            & 25\%                                                                                             \\ \hline
		Caudate(R)                       & Caudate(R)    & 0\%                                                                                       & 0\%                                                                                            & 12.5\%                                                                                             \\ \hline
		Deep Opercular Cortex(L)         & DOC(L)        & 25\%                                                                                        & 37.5\%                                                                                            & 50\%                                                                                              \\ \hline
		Deep Opercular Cortex(R)         & DOC(R)        & 25\%                                                                                        & 0\%                                                                                            & 0\%                                                                                             \\ \hline
		anterior-Middle Frontal Gyrus(L) & a-MFG(L)      & 50\%                                                                                        & 75\%                                                                                            & 75\%                                                                                             \\ \hline
		anterior-Middle Frontal Gyrus(R) & a-MFG(R)      & 25\%                                                                                       & 12.5\%                                                                                            & 37.5\%                                                                                              \\ \hline
		ventrical-Premotor Area(L)       & v-preMA(L)    & 100\%                                                                                        & 100\%                                                                                            & 100\%                                                                                              \\ \hline
		dorsal-Premotor Area(L)          & d-preMA(L)    & 62.5\%                                                                                        & 50\%                                                                                            & 50\%                                                                                             \\ \hline
		Premotor Area(R)                 & preMA(R)      & 12.5\%                                                                                       & 0\%                                                                                            & 0\%                                                                                             \\ \hline
		pre-Supplementary Motor Area     & pre-SMA       & 100\%                                                                                        & 100\%                                                                                             & 100\%                                                                                              \\ \hline
		Supra-Marginal Gyrus(L)          & SupraMG(L)    & 62.5\%                                                                                       & 50\%                                                                                             & 62.5\%                                                                                            \\ \hline
		Supra-Marginal Gyrus(R)          & SupraMG(R)    & 25\%                                                                                        & 0\%                                                                                            & 0\%                                                                                             \\ \hline
		Wernicke's Area(L)               & WA(L)         & 75\%                                                                                       & 50\%                                                                                            & 100\%                                                                                             \\ \hline
		Wernicke's Area(R)               & WA(R)         & 50\%                                                                                        & 12.5\%                                                                                            & 12.5\%                                                                                             \\ \hline
	\end{tabular}
\end{table}

\section{Link weights in individual networks}
\begin{table}[H]
		\caption[Link's weight between pairs of fROI in each individual network ]{Link's weight in individual Monolingual}
	\label{table:individual-M}
	\centering

	\begin{tabular}{|c|c|c|c|c|c|c|c|c|c|}
		\hline
		Link label & fROI pairs \#/Subjects & 1    & 2    & 3    & 4    & 5    & 6    & 7    & 8    \\ \hline
		A          & BA(L) - v-preMA(L)     & 3.94 & 3.27 & 8.11 & 1.96 & 0.01 & 1.13 & 3.03 & 4.10 \\ \hline
		B          & pre-SMA - v-preMA(L)   & 5.25 & 1.44 & 2.80 & 3.03 & 0.89 & 0.43 & 1.39 & 1.26 \\ \hline
		C          & pre-SMA - BA(L)        & 1.99 & 0.88 & 2.07 & 2.43 & 0.26 & 0.12 & 0.09 & 0.40 \\ \hline
		D          & BA(L) - WA(L)          & 0.00 & 0.19 & 1.09 & 1.75 & 0.00 & 0.01 & 0.01 & 0.02 \\ \hline
		E          & WA(L) - v-preMA(L)     & 0.00 & 0.93 & 0.01 & 1.03 & 0.00 & 0.05 & 0.01 & 0.16 \\ \hline
	\end{tabular}

\end{table}

\begin{table}[H]
	\caption[Link's weight between pairs of fROI in each individual network]{Link's weight in individual Bilingual English}
	\label{table:individual-BE}
	\centering
		\begin{tabular}{|c|c|c|c|c|c|c|c|c|c|}
			\hline
			Link label & fROI pairs \#/Subjects & 1    & 2    & 3    & 4    & 5    & 6    & 7    & 8    \\ \hline
			A          & BA(L) - v-preMA(L)     & 1.24 & 1.25 & 4.27 & 2.68 & 2.81 & 1.79 & 1.60 & 1.93 \\ \hline
			B          & pre-SMA - v-preMA(L)   & 4.24 & 0.01 & 4.88 & 1.64 & 1.78 & 0.04 & 2.70 & 1.58 \\ \hline
			C          & pre-SMA - BA(L)        & 1.49 & 0.14 & 0.74 & 0.89 & 1.64 & 0.18 & 0.01 & 0.77 \\ \hline
			D          & BA(L) - WA(L)          & 0.00 & 0.00 & 0.00 & 0.29 & 0.06 & 0.04 & 0.00 & 0.05 \\ \hline
			E          & WA(L) - v-preMA(L)     & 0.00 & 0.00 & 0.00 & 0.27 & 0.01 & 0.09 & 0.00 & 0.03 \\ \hline
		\end{tabular}
\end{table}
\begin{table}[H]
	\caption[Link's weight between pairs of fROI in each individual network]{Link's weight in individual Bilingual Spanish}
	\label{table:individual-BS}
	\centering
			\begin{tabular}{|c|c|c|c|c|c|c|c|c|c|}
				\hline
				Link label & fROI pairs \#/Subjects & 1    & 2    & 3     & 4    & 5    & 6    & 7    & 8    \\ \hline
				A          & BA(L) - v-preMA(L)     & 8.29 & 1.59 & 11.46 & 1.05 & 4.32 & 0.60 & 2.91 & 2.56 \\ \hline
				B          & pre-SMA - v-preMA(L)   & 6.61 & 0.75 & 5.21  & 2.58 & 5.62 & 1.05 & 2.18 & 4.11 \\ \hline
				C          & pre-SMA - BA(L)        & 2.68 & 0.82 & 1.53  & 0.89 & 3.54 & 0.05 & 0.29 & 1.97 \\ \hline
				D          & BA(L) - WA(L)          & 0.02 & 0.17 & 0.01  & 0.67 & 0.11 & 0.01 & 0.02 & 0.02 \\ \hline
				E          & WA(L) - v-preMA(L)     & 0.05 & 0.18 & 0.42  & 0.01 & 0.01 & 0.08 & 0.18 & 0.06 \\ \hline
			\end{tabular}
\end{table}
\section{Link weights in common networks}	
\begin{table}[H]
	\caption[Link's weight between pairs of fROI in common network ]{Link's weight in common Monolingual}
	\label{table:common-M}
	\centering
	\begin{tabular}{|c|c|c|}
		\hline
		Link label & fROI pairs & mean $\pm$ stdv \\ \hline
		A          & BA(L) - v-preMA(L)     & 3.20 $\pm$ 2.28 \\ \hline
		B          & pre-SMA - v-preMA(L)   & 2.06 $\pm$ 1.46 \\ \hline
		C          & pre-SMA - BA(L)        & 1.03 $\pm$ 0.91 \\ \hline
		D          & BA(L) - WA(L)          & 0.51 $\pm$ 0.67 \\ \hline
		E          & WA(L) - v-preMA(L)     & 0.37 $\pm$ 0.44 \\ \hline
	\end{tabular}
\end{table}

\begin{table}[H]
	\caption[Link's weight between pairs of fROI in common network ]{Link's weight in common Bilingual English}
	\label{table:common-BE}
	\centering
	\begin{tabular}{|c|c|c|}
		\hline
		Link label & fROI pairs \#/Subjects & mean $\pm$ stdv \\ \hline
		A          & BA(L) - v-preMA(L)     & 2.19 $\pm$ 0.96 \\ \hline
		B          & pre-SMA - v-preMA(L)   & 2.11 $\pm$ 1.65 \\ \hline
		C          & pre-SMA - BA(L)        & 0.73 $\pm$ 0.57 \\ \hline
		D          & BA(L) - WA(L)          & 0.11 $\pm$ 0.11 \\ \hline
		E          & WA(L) - v-preMA(L)     & 0.10 $\pm$ 0.10 \\ \hline
	\end{tabular}
\end{table}

\begin{table}[H]
	\caption[Link's weight between pairs of fROI in common network ]{Link's weight in common Bilingual Spanish}
	\label{table:common-BS}
	\centering
	\begin{tabular}{|c|c|c|}
		\hline
		Link label & fROI pairs \#/Subjects & mean $\pm$ stdv \\ \hline
		A          & BA(L) - v-preMA(L)     & 4.10 $\pm$ 3.60 \\ \hline
		B          & pre-SMA - v-preMA(L)   & 3.51 $\pm$ 2.05 \\ \hline
		C          & pre-SMA - BA(L)        & 1.47 $\pm$ 1.13 \\ \hline
		D          & BA(L) - WA(L)          & 0.13 $\pm$ 0.21 \\ \hline
		E          & WA(L) - v-preMA(L)     & 0.13 $\pm$ 0.13 \\ \hline
	\end{tabular}
\end{table}


\section{Centrality measurement}
\begin{table}[H]
	\caption[Centrality measurements]{Centrality measurements in all three groups.}
	\centering
	\begin{tabular}{|c|c|c|c|c|c|}
		\hline
		Centrality type              & Groups /fROIs     & pre-SMA & BA(L) & WA(L)  & v-preMA(L) \\ \hline
		\multirow{3}{*}{Degree}      & Monolingual       & 0.63    & 0.61  & 0.37   & 0.67       \\ \cline{2-6} 
		& Bilingual English & 0.59    & 0.55  & 0.47   & 0.67       \\ \cline{2-6} 
		& Bilingual Spanish & 0.62    & 0.52  & 0.30   & 0.64       \\ \hline
		\multirow{3}{*}{Closeness}   & Monolingual       & 1.04    & 1.05  & 0.54   & 1.08       \\ \cline{2-6} 
		& Bilingual English & 0.99    & 0.96  & 0.42   & 1.06       \\ \cline{2-6} 
		& Bilingual Spanish & 1.14    & 1.08  & 0.63   & 1.15       \\ \hline
		\multirow{3}{*}{Betweeness}  & Monolingual       & 81.14   & 63.21 & 80.00  & 59.65      \\ \cline{2-6} 
		& Bilingual English & 58.16   & 36.19 & 119.10 & 50.20      \\ \cline{2-6} 
		& Bilingual Spanish & 92.71   & 62.76 & 92.37  & 72.22      \\ \hline
	\end{tabular}
\label{table:centrality}
\end{table}


\clearpage

\end{document}